\documentclass[journal, ,onecolumn,12pt]{IEEEtran}
\usepackage{graphicx} 
\usepackage{xcolor}
\usepackage{amsmath} 
\usepackage{amsfonts}
\usepackage{algorithm}
\usepackage{algpseudocode}
\usepackage{bbm}
\usepackage{subcaption}
\usepackage{diagbox}
\usepackage{multirow}
\usepackage{algorithmicx}

\algblockdefx{MRepeat}{EndRepeat}[1]{\textbf{repeat} #1}{}
\algnotext{EndRepeat}

\begin{document}
\newcommand{\NOTE}[1]{{\color{black} #1}}
\newcommand{\NOTEADD}[1]{{\color{purple} #1}}
\newcommand{\TODO}[1]{{\color{red} TODO: #1}}
\newcommand{\TODEL}[1]{{\color{red} #1}}
\title{Analog Quantum Asynchronous Event-Based Graph Neural Network}

\author{Kristian Sotirov,
        Shaheen Acheche,
        Antonio A. Gentile, and 
        Osvaldo Simeone, \IEEEmembership{Fellow, IEEE}
\vspace{-0.5cm} 
\thanks{
Kristian Sotirov is with the King’s Communications, Learning and Information Processing (KCLIP) lab within  the Centre for Intelligent Information Processing Systems (CIIPS) at the Department of Engineering,  King’s College London, London, WC2R 2LS, UK.  
Shaheen Acheche, and Antonio A. Gentile are with Pasqal SAS, 24 Av. Emile Baudot, 91120 Palaiseau, France (email: shaheen.acheche@pasqal.com; andrea.gentile@pasqal.com). 
Osvaldo Simeone is with the Institute for Intelligent Networked Systems  (INSI),  Northeastern University London, One Portsoken Street, London E1 8PH, United Kingdom   (email: o.simeone@northeastern.edu).

The work of K. Sotirov was supported by King's College London and Pasqal through the King’s Quantum Centre for Doctoral Training. The work of O. Simeone was supported  by the European Research Council (ERC) under the European Union’s Horizon Europe Programme (grant agreement No. 101198347), by~an Open Fellowship of the EPSRC (EP/W024101/1), and~by the EPSRC project (EP/X011852/1).

Code can be found at https://github.com/ksotirov/analog-quantum-AEGNN
}
}

\maketitle

\begin{abstract}

Asynchronous, event-based graph neural networks (AEGNNs) have recently emerged as an efficient paradigm for processing the sparse and high-temporal-resolution data from event cameras. In this paper, we propose quantum analog AEGNNs (QA-AEGNNs), a novel framework to implement an AEGNN on a neutral-atom quantum computer. Neutral-atom quantum processors offer a programmable analog quantum computing platform based on controllable Rydberg-atom interactions. To this end, we map the streaming event data to an array of trapped neutral atoms, where each atom represents a graph node (event) and is positioned such that geometric proximity reflects the spatio-temporal neighborhood of events. The native Rydberg Hamiltonian of the quantum processor is programmed to mirror the message-passing computations of the AEGNN, with atomic qubit states serving as node feature embeddings and inter-atom interactions realizing graph edges. Furthermore, we propose a hybrid quantum-classical training scheme in which the analog Hamiltonian parameters (e.g., laser pulse amplitudes and detunings) are optimized using classical feedback to learn the quantum AEGNN model from data. Our approach leverages the continuous Hamiltonian dynamics and massive parallelism of neutral-atom quantum systems to natively execute event-based graph computations with potential accuracy improvements. 

\end{abstract}

\section{Introduction}\label{sec:intro}

\subsection{Context and Motivation}

The convergence of quantum computing, neuromorphic sensing, and graph neural networks presents a compelling opportunity to address fundamental challenges in event processing \cite{brand2024quantum,chen2025stochastic}. Event cameras, also known as dynamic vision sensors (DVS), represent a paradigm shift in visual sensing by asynchronously detecting per-pixel brightness changes with microsecond temporal resolution and low power consumption \cite{gallego2020event}. Unlike conventional frame-based cameras that capture images at fixed rates, event cameras generate sparse, asynchronous streams of events, each encoding the spatial location, timestamp, and polarity of a brightness change. 
This bio-inspired sensing modality offers significant advantages for high-speed motion tracking, robotics, and autonomous systems, but poses substantial challenges for classical processing architectures designed for dense, synchronous data \cite{schaefer2022aegnn}.

 \begin{figure}
	\centering
	\includegraphics[width=\textwidth]{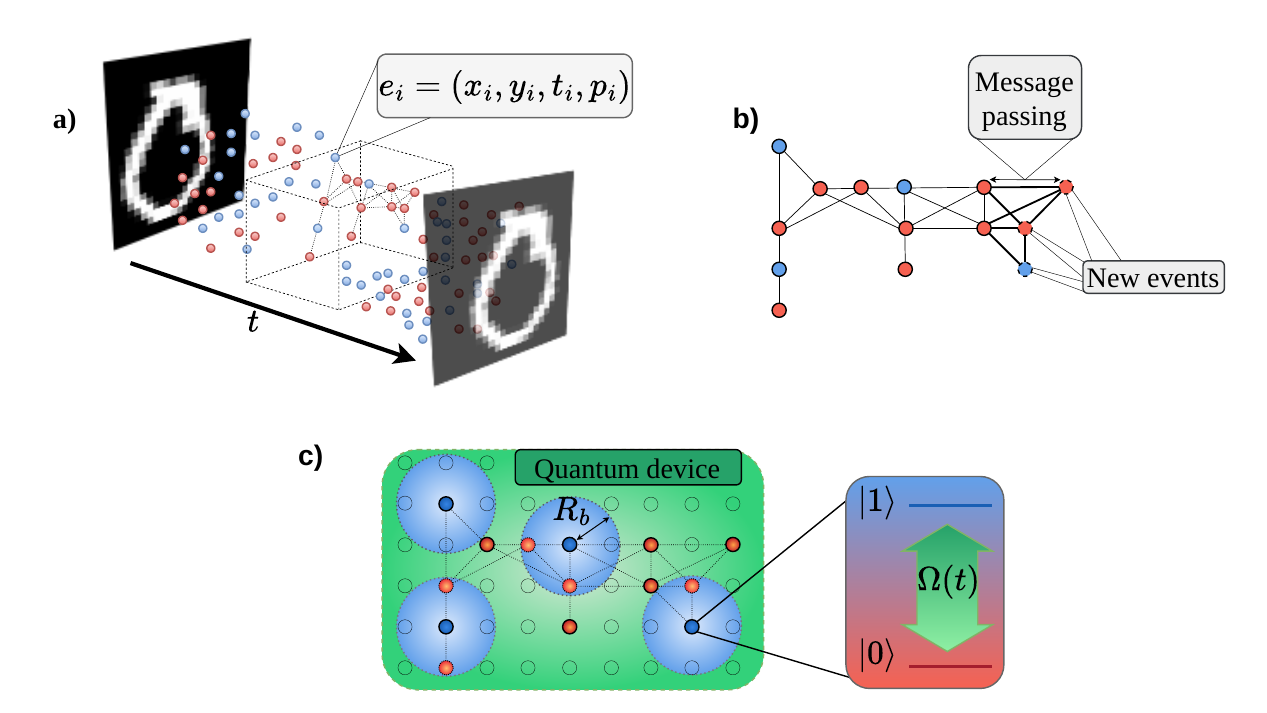}
	\caption{(a) An event-driven sensor, such as a neuromorphic camera \cite{gallego2020event}, produces a stream of events characterized by a spatial coordinate,  time stamp and a polarity. (b) In AEGNN \cite{schaefer2022aegnn}, a graph is constructed online by including a new node for each new event, with edges formed by events in close spatio-temporal proximity. After graph construction, a classical message passing graph neural network (MP-GNN) is employed. (c) In the proposed quantum analog AEGNN (QA-AEGNN), events are mapped to atoms within a neutral-atom quantum computer, and Hamiltonian evolution\NOTE{, characterized by Rabi frequency $\Omega(t)$ and Rydberg blockade radius $R_b$, } natively implements information mixing across the events. }
	\label{fig:scara}
\end{figure}

Graph neural networks (GNNs) provide a natural framework for processing event camera data by representing events as nodes in a spatio-temporal graph. Specifically, \emph{asynchronous event-based GNNs} (AEGNNs) construct dynamic graphs where nodes correspond to events and edges connect spatially and temporally proximate events, enabling message-passing algorithms to aggregate information across the event stream \cite{schaefer2022aegnn}. This is illustrated in Fig. \ref{fig:scara}, in which events produced by an event-driven camera are represented as dots in in the space of spatial and temporal coordinates. These events are mapped to nodes of a dynamic graph, which updates node features based on message passing. As event camera resolution and frame rates continue to increase, the computational demands for \NOTE{ efficient data} processing remain substantial, motivating the exploration of alternative computing paradigms.

Quantum computing offers a promising avenue for accelerating graph-based computations. Analog quantum simulation using neutral-atom arrays is particularly well suited for this task due to its scalability \cite{preskill2025beyond} and flexibility in qubit connectivity \cite{gyurik2025quantum, gentile2025hardware}.  Neutral-atom quantum processors leverage arrays of ultra-cold atoms trapped by optical tweezers, where each atom's internal states serve as qubits and inter-atomic Rydberg interactions enable programmable many-body dynamics \cite{Henriet2020quantumcomputing, wurtz2023aquila}. A key advantage of these systems is their ability to natively embed graph structures through the spatial arrangement of atoms: by positioning atoms such that their physical distances reflect desired graph connectivity, the natural Rydberg blockade phenomenon automatically implements graph adjacency constraints \cite{ebadi2022quantum}. This capability has been successfully exploited for solving combinatorial optimization problems such as Maximum Independent Set on graphs with hundreds of qubits \cite{nguyen2023quantum}.

Recent theoretical and experimental work has demonstrated the potential of quantum processors for graph machine learning tasks. Quantum feature maps for graph neural networks have been implemented on neutral-atom hardware with up to 32 qubits \cite{albrecht2023quantum}, showing competitive performance with classical methods on molecular property prediction tasks. The quantum evolution kernel framework \cite{henry2021quantum} has established that graph topology can be encoded in Hamiltonian parameters, with time evolution producing measurements that capture structural graph properties. Furthermore, large-scale demonstrations of quantum reservoir computing with over 100 qubits \cite{kornjaca2024large} have validated the scalability of analog quantum approaches for machine learning, achieving performance competitive with classical methods while avoiding the barren plateau problems that plague gradient-based quantum circuit optimization.

The intersection of these two domains remains largely unexplored, despite these advances in quantum graph learning and event-based neural networks. No prior work has investigated quantum implementations of asynchronous event-based graph neural networks, representing a significant gap given the natural synergies between these approaches. The temporal dynamics inherent in event streams naturally map to the continuous evolution of quantum systems under time-dependent Hamiltonians. The directed, causal structure of event graphs aligns well with the unitary evolution of quantum states. Moreover, the quantum superposition enables coherent evolution over exponentially many basis states, while the native graph embedding capabilities of neutral-atom arrays offer potential advantages for processing high-throughput event streams with reduced latency and energy consumption.

This work aims to bridge this gap by proposing \emph{quantum analog AEGNNs} (QA-AEGNNs), a novel framework that leverages the native Hamiltonian dynamics of neutral-atom quantum processors to implement asynchronous event-based graph neural networks. By mapping event streams to configurable atomic arrays and programming the Rydberg Hamiltonian to mirror classical message-passing operations, we demonstrate how analog quantum evolution can perform graph-based event processing.

\subsection{Related Works}

Our work draws upon three interconnected research areas: quantum graph neural networks, event-based neural network architectures, and analog quantum computing for machine learning. We review the key developments in each domain and identify how they inform our proposed approach.

\subsubsection{Quantum Graph Neural Networks}

The theoretical foundations for quantum graph neural networks were established by reference  \cite{verdon2019quantum}, which proposed quantum circuit architectures for graph-structured data where parameterized unitaries act on node features and entangling gates follow graph edges. This work inspired extensive research on quantum approaches to graph learning, with recent surveys \cite{ceschini2024quantum} categorizing QGNNs into convolutional, recurrent, spatial-temporal, and equivariant variants.

For neutral-atom implementations specifically, the quantum evolution kernel framework  \cite{henry2021quantum} provides a gradient-free approach where graph topology is encoded in Hamiltonian parameters and quantum time evolution produces feature maps that preserve graph structure. This method is particularly well-suited to analog quantum processors, as it leverages natural many-body dynamics rather than requiring compilation into discrete gate sequences. The reference \cite{albrecht2023quantum} provided the first experimental validation of this approach on Pasqal's neutral-atom quantum processor, demonstrating graph classification on molecular datasets with up to 32 qubits.

A comprehensive analysis of graph algorithms on neutral-atom quantum processors \cite{EPJA2024graph} establishes that the reconfigurable geometry of atom arrays enables native graph embedding at the hardware level, with Rydberg blockade naturally implementing adjacency constraints. This capability has been exploited for combinatorial optimization on graphs with hundreds of nodes \cite{ebadi2022quantum,nguyen2023quantum}, validating the scalability of neutral-atom platforms for graph-structured problems. Recent work  \cite{simard2025learning} demonstrated the reverse direction, using classical graph neural networks to learn Rydberg Hamiltonians from experimental data, further confirming the bidirectional compatibility between GNN architectures and Rydberg quantum systems.

Theoretical work on quantum neural network trainability has identified permutation-equivariant architectures as a solution to barren plateau problems \cite{schatzki2024theory}. By preserving graph symmetries through hardware-native constraints, these architectures avoid vanishing gradients in high-dimensional parameter spaces and generalize well from limited training data—properties critical for practical quantum machine learning applications.

\subsubsection{Event-Based Graph Neural Networks and Neuromorphic Processing}

The foundation for event-based graph neural networks was established by  \cite{schaefer2022aegnn}, who introduced the AEGNN architecture for processing event camera data. Their key insight was to represent asynchronous event streams as dynamic spatio-temporal graphs, where events become nodes characterized by position $(x,y,\beta t)$ and polarity, with edges connecting events within a defined spatio-temporal radius. By restricting message-passing updates to nodes affected by new events, AEGNNs achieve significant computational savings \cite{gehrig2024recurrent}.

Hardware implementations have rapidly followed these algorithmic advances. Reference \cite{yang2024evgnn} developed EvGNN, the first event-driven GNN hardware acceleration on FPGA platforms, and the work  \cite{jeziorek2024embedded} demonstrated embedded graph convolutional networks on  FPGAs.

Spiking graph neural networks represent a parallel development with direct relevance to our quantum approach. These architectures combine the event-driven nature of spiking neural networks with graph-structured message passing, achieving further energy efficiency through sparse, binary activations \cite{zhao2024drsgnn,yang2024manifold,simeone2026modern}. The temporal dynamics of spiking neurons naturally align with the asynchronous processing of event streams, offering a biological precedent for time-continuous neural computation on graphs.

\subsubsection{Analog Quantum Computing for Machine Learning}

Analog quantum computing leverages the continuous Hamiltonian evolution of quantum many-body systems for computation, offering an alternative to gate-based digital quantum circuits. For neutral-atom platforms specifically, the programmable Rydberg Hamiltonian enables direct implementation of many-body interactions without requiring decomposition into two-qubit gates, potentially offering advantages in circuit depth and coherence time utilization \cite{Henriet2020quantumcomputing,gentile2025hardware,weiss2017quantum}.  \NOTE{It is also noted that the Hamiltonian dynamics underlying neutral atom quantum computing platforms can be efficiently emulated within trapped-ion architectures, enabling equivalent many-body quantum evolutions across these distinct physical systems \cite{blatt2012quantum}.}

Quantum spiking neural networks provide a direct bridge between neuromorphic computing and quantum systems. The quantum leaky integrate-and-fire (QLIF) neuron \cite{brand2024quantum} requires only two rotation gates without CNOT operations, enabling construction of quantum spiking networks trained on neuromorphic datasets with faster convergence than variational quantum circuits. More recently, stochastic quantum spiking networks \cite{chen2025stochastic} have introduced event-driven probabilistic spike generation with internal quantum memory and local learning rules, combining the temporal efficiency of neuromorphic computing with exponentially large quantum state spaces.

\subsection{Main Contributions}

This paper makes the following key contributions:

\begin{enumerate}
\item \textbf{Quantum analog AEGNN framework}: We introduce QA-AEGNN, a novel architecture that maps asynchronous event-based graph neural networks to neutral-atom quantum processors. Our framework encodes event streams as configurable atomic arrays where spatial proximity reflects spatio-temporal event relationships, and programs the Rydberg Hamiltonian to implement graph message-passing operations through natural many-body interactions. QA-AEGNN embeds dynamic event graphs onto quantum hardware using either 3D atomic arrangements (direct coordinate mapping) and 2D arrays, enabling adaptation to current experimental capabilities. We evaluate two different methods to encode binary node features, either through state initialization or through the modulation of local detuning coefficients  \cite{kornjaca2024large}.

\item \textbf{Hybrid quantum-classical lightweight training scheme}: We propose a lightweight variational  approach for the data-driven optimization  of QA-AEGNN, whereby the duration of the quantum time evolution is jointly trained with the parameters of the final classical classifier. This way, the quantum model effectively acts as a form of data-driven reservoir model with  internal  structure adapted to the input data  \cite{kornjaca2024large,gyurik2025quantum}.

\item \textbf{Experimental Validation}: We present  results comparing QA-AEGNN against classical AEGNN on synthetic graph classification tasks and neuromorphic datasets, demonstrating competitive or superior performance in distinguishing graphs with subtle structural differences.
\end{enumerate}

 The remainder of this paper is organized as follows. Sec. \ref{sec:aegnn} describes the  AEGNN framework \cite{schaefer2022aegnn}, and the specific implementation adopted in this work. Sec. \ref{sec:neutral_atoms} outlines the Hamiltonian dynamics employed by neutral-atom quantum computers as necessary background information. Sec. \ref{sec:qa-aegnn} presents the QA-AEGNN model, including implementation details. We examine the performance of the QA-AEGNN in Sec. \ref{sec:experiments}, presenting numerical results, followed by conclusions in Sec. \ref{sec:discussion}.

\section{Classical Asynchronous Event-Based Graph Neural Networks}\label{sec:aegnn}

The AEGNN model was introduced in \cite{schaefer2022aegnn} to process the evolving graph of events produced by event-based neuromorphic cameras. This section reviews the classical model to facilitate the introduction of the proposed quantum model in subsequent sections.

\subsection{Dynamic Event-Driven Graph Construction}

Each event $e_i$ is a tuple $e_i=(x_i, y_i, t_i, p_i)$ encoding the pixel coordinate $(x_i,y_i)$ where a brightness change occurred, the timestamp $t_i$, and the polarity $p_i \in \{+1,-1\}$ indicating the sign of the brightness change. AEGNN treats each event as a graph node and connects edges between nodes that are close in space and time, maintaining a sliding window of events. 

\begin{figure}
    \centering
    \includegraphics[width=0.7\textwidth]{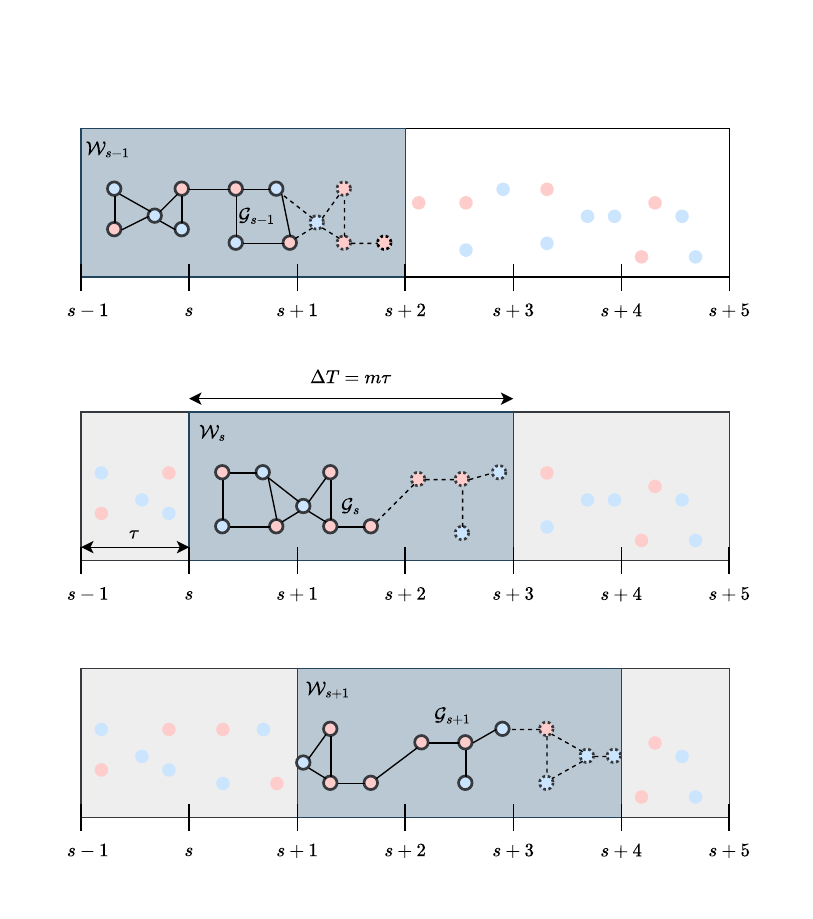}
    \caption{Illustration of the time windows $\mathcal{W}_{s-1}$, $\mathcal{W}_s$, and $\mathcal{W}_{s+1}$ of duration $\Delta T = m\tau$, used to process event streams produced by a neuromorphic camera. Legacy nodes and edges (present in the previous window) are indicated with continuous lines, while new nodes and edges are indicated with dotted lines. The color denotes the polarity $p_i \in \{+1, -1\}$, with blue indicating $p_i = +1$ and red indicating $p_i = -1$.} 
    \label{fig:aegnn_diagram}
\end{figure}

Formally, as illustrated in Fig. \ref{fig:aegnn_diagram}, we fix a time window duration $\Delta T$ and an inter-window shift $\tau$ with $\Delta T = m\tau$ for some integer $m$. AEGNN operates over a discrete time index $s = 1, 2, \dots, S$, where $s$ indicates the time window $\mathcal{W}_s = [s\tau - \Delta T, s\tau ]$. An undirected event graph $\mathcal{G}_s =(\mathcal{V}_s,\mathcal{E}_s)$ is constructed for each window $\mathcal{W}_s$, which includes one node $i\in \mathcal{V}_s$ for each event in window $\mathcal{W}_s$. Accordingly, the graph has one node for all events $e_i$ with $t_i \in \mathcal{W}_s$.

To construct the graph $\mathcal{G}_s$, the node $i$ corresponding to event $e_i$ is placed at a point $X_i=(x_i, y_i, \hat{t}_i)$ in a 3D spatio-temporal space, where $\hat{t}_i = \beta t_i$ is a scaled time coordinate for some constant $\beta > 0$. An edge $(i,j)\in \mathcal{E}_s$ is created if the distance between two event nodes in this space is below a threshold $R > 0$:
\begin{equation} \label{eq:threshold} 
\|X_i - X_j\| \le R. 
\end{equation} 
In practice, the parameter $\beta$ is chosen to normalize time differences to a range comparable to spatial distances, and the scalar $R$ defines a neighborhood radius in space-time within which events are considered adjacent. This yields a \emph{sparse} graph where each event connects only to other events near it in both location and time, capturing the natural locality of event camera data. 

\subsection{Message Passing}

Each node $i \in \mathcal{V}_s$ is assigned an initial value for the node feature $h_{s,i}^0$ when processing the subset of events \begin{equation} \label{eq:eventwins} E_s =\{ e_i : e_i = (x_i, y_i, t_i, p_i)\in E, t_i \in \mathcal{W}_s\} \end{equation} in window $\mathcal{W}_s$. To determine the initial values $\{h_{s,i}^0\}_{i \in \mathcal{V}_s}$, we distinguish two types of events in a window $\mathcal{W}_s$: \textit{new events} with $e_i \notin \mathcal{W}_{s-1}$ and \textit{legacy events} with $e_i \in \mathcal{W}_{s-1} \cap \mathcal{W}_{s}$. Legacy events were present also in the previous window $\mathcal{W}_{s-1}$, enabling incremental graph construction by updating graph $\mathcal{G}_{s-1}$ as detailed in Algorithm \ref{algo:update_graph}. 

\begin{algorithm}
    \caption{Update Graph Function}
    \label{algo:update_graph}
    \textbf{Input}: Events $e_i = (x_i, y_i, t_i, p_i)$ with $t_i \in \mathcal{W}_s$; previous graph $\mathcal{G}_{s-1} = (\mathcal{V}_{s-1}, \mathcal{E}_{s-1})$; radius $R$; constant $\beta \geq 0$.
    \begin{algorithmic}[1]
        \State $\mathcal{V}_s \leftarrow \mathcal{V}_{s-1} \setminus \{i \in \mathcal{V}_{s-1}: t_i \notin \mathcal{W}_{s}\}$ \Comment{Drop nodes for events not in window $\mathcal{W}_s$}
        \State $\mathcal{E}_s \leftarrow \mathcal{E}_{s-1} \setminus \{(i,j) \in \mathcal{E}_{s-1}: i \notin \mathcal{V}_{s} \ \text{or} \ j \notin \mathcal{V}_{s} \}$ \Comment{Drop edges}
        \State $\mathcal{V}_s \leftarrow \mathcal{V}_s \cup \{i : e_i = (x_i, y_i, \beta t_i, p_i) $ with $t_i \in \mathcal{W}_{s} \setminus \mathcal{W}_{s-1}\} $\Comment{Add nodes for new events}
        \State $\mathcal{E}_s \leftarrow \mathcal{E}_s \cup \{(i,j):  i, j \in \mathcal{V}_s,\ \text{and} \ ||X_i - X_j|| \leq R\}$ \Comment{Add edges}
    \end{algorithmic}
    \hspace*{\algorithmicindent}\textbf{Return:} Graph $\mathcal{G}_s = (\mathcal{V}_s, \mathcal{E}_s)$
\end{algorithm}

For new events, the polarity $p_i$ of event $e_i$ is used as the initial value for the corresponding node feature: $h_{s,i}^0 = p_i$. For legacy events $e_i$, the initial value equals the final node feature from processing the previous window. Denoting the latter as $h_{s-1,i}^L$, we have $h_{s,i}^0 = h_{s-1,i}^L$ for legacy nodes. This initialization scheme enables temporal continuity across sliding windows. In this work, we do not consider edge features, which could include relative spatial or temporal displacements between events \cite{schaefer2022aegnn}. 

Once the graph $\mathcal{G}_s = (\mathcal{V}_s,\mathcal{E}_s)$ is constructed using Algorithm \ref{algo:update_graph} and the node features $\{h_{s,i}^0\}$ are initialized, a message-passing graph neural network (MP-GNN) propagates information across this graph to compute a sequence of updated node features $h_{s,i}^1, h_{s,i}^2,\dots,h_{s,i}^L$ for each node $i \in \mathcal{V}_s$ across $L$ layers. \NOTE{The MP-GNN framework is outlined in Algorithm \ref{algo:AEGNN}, and detailed next.}

Mathematically, each $l$-th MP-GNN layer is defined by a \textit{permutation-invariant} aggregation function and a combination function. In this work, in order to ensure consistency with practical quantum models based on reservoir computing \cite{henry2021quantum}, the message passing model follows fixed dynamics across layers $l = 1,\dots,L$ that are described by the standard sum-aggregation function \cite{xuHowPowerfulAre2019} 

\begin{equation}\label{eq:aggregate}
a_{s,i}^{l} = \sum_{j \in \mathcal{N}_{s, i}} {h}_{s,j}^{l-1},
\end{equation}
where $\mathcal{N}_{s, i} = \{j | (i,j) \in \mathcal{E}_s\}$ denotes the neighbors of node $i$ in graph $\mathcal{G}_s$, while the combination function is given by a non-linearity of the form \cite{hamilton2017inductive}
\begin{equation}\label{eq:combine}
\sigma_{\alpha_s} \left(h_{s,i}^{l-1} + a_{s,i}^{l} \right).
\end{equation}

In (\ref{eq:combine}), the notation $\sigma_{\alpha_s}(\cdot)$ denotes a non-decreasing function, whose shape can be controlled via a single tunable parameter $\alpha_s$. This way, the model dynamics across layers are fully specified by the scalar parameters $\{\alpha_s\}_{s=1}^S$. As a specific instance of the combination function (\ref{eq:combine}) we will adopt the standard sigmoid with control parameter $\alpha_s$, i.e., 
\begin{equation}
    \sigma_{\alpha_s}(x) = \frac{1}{1 + e^{\alpha_sx}}.
\end{equation}
Sec. \ref{sec:aegnn_training} will discuss the optimization of parameters $\{\alpha_s\}_{s = 1}^S$.

\subsection{Graph Classification}

We focus on a graph classification task in which the features extracted by the AEGNN are passed as input to a classifier model $f_s(\cdot)$, such as a support vector machine (SVM) or a neural network, which produces a probability distribution over $K$ classes. We specifically leverage \textit{histogram pooling} \cite{shervashidzeWeisfeilerLehmanGraphKernels2011}, applied separately to each layer $l$.  The classifier $f_s(\cdot)$ generally depends on the window index $s$. 

Accordingly, after applying the aggregation function (\ref{eq:aggregate}) and the combine function (\ref{eq:combine}) on all nodes, for each layer $l$, we quantize all node features $h_{s,i}^l$ using the function 
\begin{equation}
    \hat{h}_{s,i}^l = \text{Q}_{B, a^l,b^l}(h_{s,i}^l),
\end{equation} 
which clips the input value within the interval $[a^l,b^l]$ and  applies uniform quantization over $B$ bins in the same interval. The interval $[a^l,b^l]$ is determined to cover a sufficiently large mass of the empirical distribution of the features $h_{s,i}^l$, e.g., the interval between the 10th and 90th empirical percentiles. The quantized features $\{\hat{h}_{s,i}^l\}_{i\in \mathcal{V}}$ are aggregated into a $B\times 1$ histogram $z_s^l = \text{Hist}(\{\hat{h}_{s, i}^l\}_{i \in \mathcal{V}_s})$, with each $j$-th entry of vector $z_s^l$ counting the number of nodes in $\mathcal{V}_s$ with feature $\hat{h}_{s,i}^l$ equal to $j$. 

The resulting $L$ histograms $\{z_s^l\}_{l=1}^L$ are concatenated into the $LB\times1$ vector 
\begin{equation}
    z_s = [(z_s^1)^T,(z_s^2)^T, \dots, (z_s^L)^T]^T
\end{equation}
to serve as input to the classifier model $f_s(\cdot)$.

In this way, the model makes a decision $f_s(z_s)$ at the end of each \NOTE{window $\mathcal{W}_s$}. In this study, we consider the majority rule as a strategy to aggregate the class predictions. Accordingly, denoting as $\hat{k}_s$ the output of the classifier at \NOTE{window $\mathcal{W}_s$}, i.e., $\hat{k}_s = f_s(z_s)$, the final classification decision $\hat{k}$ is given by 
    \begin{equation}\label{eq:majority_rule}
        \hat{k} = \arg \max_{\kappa} \sum_{s = 1}^S \mathbbm{1}(\hat{k}_s = \kappa),
    \end{equation} 
    where $\mathbbm{1}(\cdot)$ is the indicator function.

Algorithm \ref{algo:AEGNN} summarizes the complete AEGNN processing pipeline.

\begin{algorithm}
    \caption{AEGNN}
    \label{algo:AEGNN}
    \textbf{Input}: Event sequence $E = (e_1, e_2, \dots, e_N)$, with event $e_i = (x_i, y_i,t_i, p_i)$ where $t_1 \leq t_2 \leq \dots \leq t_N$ and polarity $p_i \in \{-1, +1\}$; radius $R$; time window duration $\Delta T$; window shift $\tau$ where $\Delta T = m\tau$ for some integer $m$;  set of parameters $\{\alpha_s\}_{s=1}^S$ ; constant $\beta \geq 0$; number of bins $B$; number of layers $L$; clipping boundaries $(a^1, b^1), (a^2, b^2), \dots,(a^L,b^L)$, with $a^l < b^l, \forall l \in \{1,2,\dots, L\}$.  
    \begin{algorithmic}[1]
        \State $\mathcal{V}_{0} \leftarrow \emptyset$ and $\mathcal{E}_{0} \leftarrow \emptyset$ \Comment{Initialize node and edge sets}
        \State $\mathcal{G}_{0} = (\mathcal{V}_{0},\mathcal{E}_{0})$ \Comment{Initialize graph}
        \For {$s = 1,2 \dots, S$}
            \State $\mathcal{W}_s \leftarrow [s\tau - \Delta T, s\tau ]$ \Comment{Form time window}
            \State $E_s \leftarrow \{ e_i : e_i = (x_i, y_i, t_i, p_i)\in E, t_i \in \mathcal{W}_s\}$ \Comment{Create event subset}
            \State $\mathcal{G}_s = (\mathcal{V}_s,\mathcal{E}_s) \leftarrow \text{Update-Graph}(E_s, \mathcal{G}_{s-1}, R, \beta)$ \Comment{Update graph using Algorithm 1}
            \For{$e_i \in E_s$}
            \If{$e_i \in E_{s-1}$}
                \State $h_{s,i}^0 \leftarrow h_{s-1,i}^L$ \Comment{Legacy nodes: use previous feature}
            \Else
                \State $h_{s,i}^0 \leftarrow p_i$ \Comment{New nodes: use polarity}
            \EndIf
            \EndFor
            \For {$l = 1,2, \dots, L$} \label{algo:lines:MPstart}
                \State $\{a_{s,i}^{l}\}_{i \in \mathcal{V}_s} \leftarrow \left \{\sum_{j \in N_{s,i}} h_{s,j}^{l-1} : i \in \mathcal{V}_s \right \}$\label{algo:lines:aggregate} \Comment{Aggregate}
                \State $\{h_{s,i}^{l}\}_{i \in \mathcal{V}_s} \leftarrow \left \{ \sigma_{\alpha_s} \left(h_{s,i}^{l-1} + a_{s,i}^{l} \right) : i \in \mathcal{V}_s \right \} $\label{algo:lines:combine} \Comment{Combine}
                \State $\{\hat{h}_{s, i}^l\}_{i \in \mathcal{V}_s} \leftarrow \{\text{Q}_{B, a^l,b^l}(h_{s, i}^l ) : {i \in \mathcal{V}_s}\}$ \Comment{Quantize}
                \State $z_s^l \leftarrow \text{Hist}(\{\hat{h}_{s, i}^l\}_{i \in \mathcal{V}_s})$ \Comment{Create histogram}
            \EndFor \label{algo:line:MPend}
            \State $z_s \leftarrow [(z_s^1)^T,(z_s^2)^T, \dots, (z_s^L)^T]^T$ \Comment{Concatenate histograms}
            \State \NOTE{$\hat{k}_s \leftarrow  f_s(z_s)$ \Comment{Class prediction at \NOTE{window $\mathcal{W}_s$}}}
            \EndFor
    \end{algorithmic}
     \hspace*{\algorithmicindent}\textbf{Return:} \NOTE{Classification decision $\hat{k} = \arg \max_{\kappa} \sum_{s = 1}^S \mathbbm{1}(\hat{k}_s = \kappa)$}
\end{algorithm}

\subsection{Training AEGNN}\label{sec:aegnn_training}

As described in the previous sections, the learnable parameters of the AEGNN algorithm encompass: (\emph{i}) the parameters $\{\alpha\}_{s = 1}^S$ in the non-linearity (\ref{eq:combine}) applied by message passing; and (\emph{ii}) the learnable parameters $\theta_s$ of the classifier models $\{f_s(\cdot)\}_{s=1}^S$. We optimize the parameters $(\alpha_s, \theta_s)$ separately for each \NOTE{window $\mathcal{W}_s$} by solving an empirical risk minimization problem based on training data.

To elaborate, denote as $\mathcal{D}$ a dataset consisting of pairs $(E, k)$, where $E = (e_1, \dots,e_N)$ is a sequence of events and $k\in\{1,...,K\}$ is a class label.  For each \NOTE{window $\mathcal{W}_s$}, consider the subset of events  $E_s$ in (\ref{eq:eventwins}), based on which the AEGNN returns a feature vector $z_s$ as detailed in Algorithm 2. We denote the features as  $z_s (E_s | \alpha_s)$ to make explicit the dependence on input $E_s$ and non-linearity parameter $\alpha_s$.  Using these features,  given classifier parameters $\theta_s$, the classifier returns the output $f_s(z_s(E_s | \alpha_s) | \theta_s)$ in every \NOTE{window $\mathcal{W}_s$}, where we have indicated explicitly the dependence on parameters $\theta_s$.

We optimize parameters $(\alpha_s, \theta_s)$ by addressing the empirical risk minimization problem: 
\begin{equation}\label{eq:loss}
\min_{(\alpha_s, \theta_s)} \mathbb{E}_\mathcal{D}[\ell(k, f_s(z_s(E_s|\alpha_s) | \theta_s))],
\end{equation}
where $\ell(\cdot,\cdot)$ is a classification loss function, such as the hinge loss or the log-loss \cite{simeone2022machine}, and $\mathbb{E}_{\mathcal{D}}[\cdot]$ represents the empirical average over dataset $\mathcal{D}$. In practice, we address problem (\ref{eq:loss}) by discretizing the set of possible values of parameter $\alpha_s$, and by solving problem (\ref{eq:loss}) over parameters $\theta_s$ separately for each value of parameter  $\alpha_s$.

\section{Neutral-Atom Analog Quantum Computers}\label{sec:neutral_atoms}

Neutral-atom quantum processors represent quantum bits using ultra-cold atoms (typically rubidium or cesium) trapped by optical tweezers in a vacuum chamber. \NOTE{Two different electronic levels of each atom are used to encode a qubit, where the computational basis vector $|0\rangle$ is expressed via the ground state, and the computational basis vector $|1\rangle$ is expressed via a Rydberg state.} Atoms in their Rydberg state exhibit strong van der Waals interactions with each other, with an interaction energy that decays as $1/r^6$ with distance $r$ between atoms.

The effective many-body Hamiltonian governing the atoms can be written as a sum of three terms: two terms affecting each qubit separately, i.e., a global \emph{drive} term and a local \emph{detuning} term, and a term defining \emph{interactions} between qubits closer than the Rydberg radius (see, e.g., \cite{Henriet2020quantumcomputing,wurtz2023aquila,weiss2017quantum}). For $N_{\text{qubit}}$ atoms, and thus $N_{\text{qubit}}$ qubits, it is
\begin{equation}\label{eq:Rydberg}
H(t) = \sum_{i=1}^{N_{\text{qubit}}} \frac{\Omega(t)}{2} \left( e^{i \phi(t)} |1_i\rangle \langle 0_i|  + e^{-i \phi(t)} |0_i\rangle \langle 1_i| \right) 
\; - \; \sum_{i=1}^{N_{\text{qubit}}} \delta_i(t)n_i 
\; + \; \sum_{i < j} \frac{C_6}{\| {r}_i - {r}_j \|^6} n_i n_j,
\end{equation}
where: 
\begin{itemize}
  \item $\Omega(t)$ is the global Rabi frequency (laser drive strength);
  \item $\phi(t)$ is the laser phase;
  \item $\delta_i(t)$ is the laser detuning, which can be positive or negative;
  \item the Rydberg state population operator is 
  \begin{equation} 
  n_i = |1_i\rangle \langle 1_i|; 
  \end{equation}
  \item ${r}_i$ is the position vector of atom $i$ in 2D or 3D, depending on the implementation;
  \item $C_6$ is the van der Waals interaction coefficient.
\end{itemize}

The (global) Rabi frequency $\Omega(t)$ coherently drives all atoms, and $\phi(t)$ is the corresponding laser phase. The first term drives transitions between states $|0\rangle$ and $|1\rangle$ on each atom at frequency $\Omega(t)$ (Rabi oscillations). 
 
The parameter $\delta_i(t)$ is a controllable detuning of the driving laser frequency from the atom's resonance.  This biases each qubit's energy, with positive detuning making the excited state energetically favorable, while negative detuning resists excitation. The detuning frequency can be split into two terms \begin{equation}\label{eq:detuning}
    \delta_i(t) = \delta^{\text{global}}(t) + \mu_i\delta^{\text{local}}(t),
\end{equation} 
where $\delta^{\text{global}}(t)$ is the global detuning term, applied uniformly on all atoms, and $\delta^{\text{local}}(t)$ is the site-dependent term, which allows individual atoms to have different local detunings, regulated by site modulation $\mu_i \in [0, 1]$.  

Finally, the last term accounts for the pairwise van der Waals interaction strength between atoms $i$ and $j$, where  $\|r_i - r_j\|$ is the distance between the two atoms and $C_6$ is a constant that depends on the Rydberg level of the atoms in the system \cite{bermot2025local}. We write 
\begin{equation}\label{eq:Vij}
V_{ij}=\frac{C_6}{\| {r}_i - {r}_j \|^6}
\end{equation}
for the energy cost accrued whenever both atoms $i$ and $j$ are simultaneously in the Rydberg state $|1\rangle$. For atoms far apart, the term $V_{ij}$ is negligible, but for nearby atoms it can be very large. The distance at which interactions dominate is called the \emph{Rydberg blockade radius} $R_b$. In practice, if two atoms are closer than $R_b$, the system dynamics strongly suppress any state where both are excited. For a constant Rabi frequency $\Omega(t) = \Omega$, and a constant detuning term $\delta(t) = \delta$, the Rydberg blockade radius is defined by the formula 

\begin{equation}\label{eq:blockade}
    R_b = \left ( \frac{C_6}{\hbar (\Omega  + \delta)}\right )^\frac{1}{6}.
\end{equation} 
 
Overall, the parameters of the Rydberg Hamiltonian are the positions $\{r_i\}_{i=1}^{N_{\text{qubit}}}$ of each atom in the initial trap configuration, and the time-dependent control waveforms for $\Omega(t)$ and $\delta_i(t)$, with the phase commonly set to $\phi(t) = 0$ \cite{lu2024digital}.

With these parameters, the device initializes all atoms (usually in state $|0\rangle$), applies the global laser pulses according to the given  schedule of the signals $\Omega(t)$ and $ \delta_i(t)$, thus evolving the system under Hamiltonian $H(t)$, and finally measures each atom. The measurement yields a bit string of length $N_{\text{qubit}}$, \NOTE{with a value 1 assigned to each atom measured in the excited state $|1\rangle$, and 0 to each atom measured in the ground state $|0\rangle$}. By repeating the experiment many times and/or running multiple experiments in parallel, one can estimate probabilities or expectation values of various observables.

\section{Quantum Analog AEGNN on a Rydberg Atom Array}\label{sec:qa-aegnn}

This paper proposes \emph{quantum analog AEGNN} (QA-AEGNN), a novel implementation of AEGNN models that leverages the native Rydberg Hamiltonian (\ref{eq:Rydberg}) of the quantum processor to mirror the message-passing computations (\ref{eq:aggregate})-(\ref{eq:combine}) of the AEGNN. In the proposed approach, atomic qubit states serve as node feature embeddings, and inter-atom interactions realize graph edges. 

\begin{figure}
    \centering
    \includegraphics[width=0.7\textwidth]{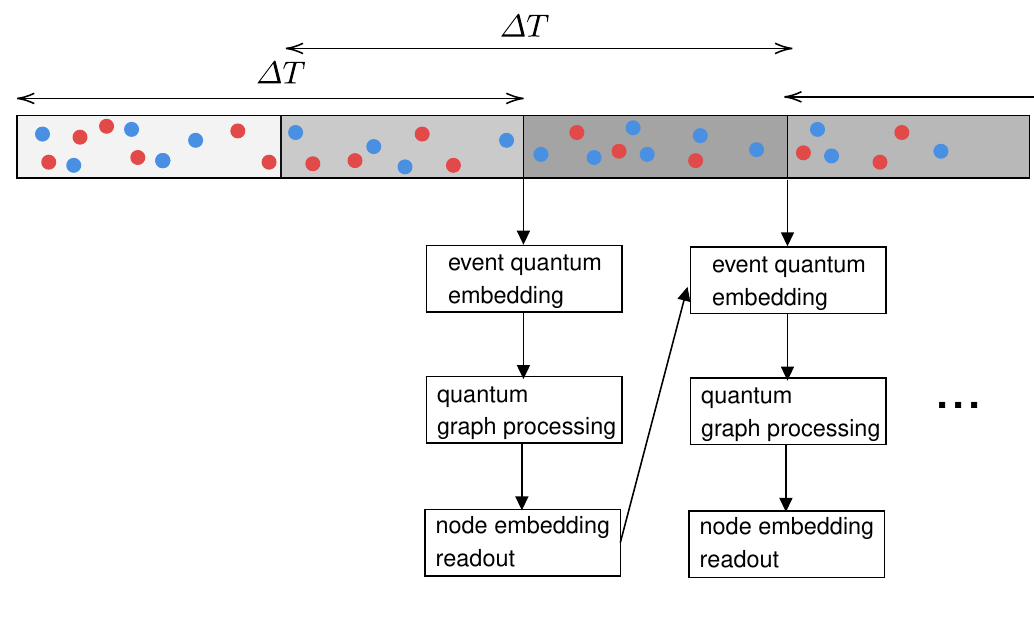}
    \caption{Illustration of the timeline of the proposed QA-AEGNN. Inside each window with duration $\Delta T$, \emph{event quantum embedding} translates new and legacy events into the neutral atom quantum hardware, followed by \emph{quantum graph processing}, where the state of the atoms is altered according to the time evolution, specified by the Hamiltonian in (\ref{eq:Rydberg}), followed by \emph{node embedding readout}, extracting the histograms.} 
    \label{fig:dia}
\end{figure}

As illustrated in Fig. \ref{fig:dia}, QA-AEGNN slides a window through the events recorded by the neuromorphic sensor over a discrete time index $s = 1,2,\dots, S$. The duration of the window is denoted as $\Delta T$, and we consider as default an implementation in which successive windows are shifted by $\Delta T/2$ (see Fig. \ref{fig:dia}). At the end of each window \NOTE{$\mathcal{W}_s$}, prior to the start of the following window \NOTE{$\mathcal{W}_{s+1}$}, the following steps are carried out:
\begin{enumerate}
    \item \emph{Event quantum embedding}: Encode the event graph captured in the current window, as well as feedback from the previous window, into the atom array;
    \item \emph{Quantum graph processing}: Perform the GNN computations via the Rydberg Hamiltonian (\ref{eq:Rydberg});
    \item \emph{Node embedding readout}: Extract measurement results at the end of each sliding window. 
\end{enumerate}

\subsection{Event Quantum Embedding: Mapping Event Data to Atomic Qubits}

The first step is to embed the event graph corresponding to the events in the current window into the quantum hardware's layout. By the design illustrated in Fig. \ref{fig:dia}, each window \NOTE{$\mathcal{W}_s$} has a number of events in common with the previous window \NOTE{$\mathcal{W}_{s-1}$}, i.e., \emph{legacy events}, as well as a number of \emph{new events}.

Denote as $\{e_i=(X_i,p_i)\}_{i=1}^{N_{\text{qubit}}}$ the events -- legacy and new -- in the current window, with $X_i=(x_i,y_i,\hat{t}_i)$. Note that the total number of events, $N_{\text{qubit}}$, generally varies from window to window. We assign each event $e_i$ to a distinct atom (qubit) in the quantum processor. We now discuss how to choose the positions $\{r_i\}_{i=1}^{N_{\text{qubit}}}$ as well as different methods to encode the node features.
 
\noindent \underline{\emph{Physical arrangement of the atoms}}: The physical position $r_i$ of each atom $i$ must be chosen \NOTE{so that neighboring graph elements in AEGNN are mapped to spatially close atoms in the quantum processor.}
\begin{itemize}

\item \emph{Neighboring nodes in the AEGNN}: \NOTE{ If two events $e_i$ and $e_j$ form an edge in the AEGNN algorithm, satisfying condition \eqref{eq:threshold}, then their corresponding atoms are placed within a distance smaller than the Rydberg blockade radius $R_b$.}

\item \emph{Non-neighboring nodes in the AEGNN}: If two events are not connected in the AEGNN, i.e., condition (\ref{eq:threshold}) is not satisfied, their atoms should be placed at a distance larger than the Rydberg blockade radius, so that the interaction term $V_{ij}$ in (\ref{eq:Vij}) is negligible.
\end{itemize}  

To satisfy these conditions, there are two possible approaches:  
\begin{itemize} 
\item \emph{3D atomic array}: If a fully 3D placement of atoms were possible, one could simply use the physical coordinates $r_i= c \cdot (x_i, y_i,\hat{t}_i)$ for some scaling constant $c$ to map the event's 3D location to the atom's 3D coordinates. 

\item \emph{2D atomic array}: In the more common case of systems that allow only a 2D placement, we set 
\begin{equation}\label{eq:classicalemb}
    r_i= M (x_i, y_i,\hat{t}_i)^T
\end{equation} 
for some $2\times 3$ matrix $M$. The default choice in Section \ref{sec:experiments} will be the matrix 

\begin{equation}\label{eq:linear_transformation}
    M = \begin{pmatrix}
        c_{xy} & 0 & c_{t}\\
        0& c_{xy} & c_{t}
    \end{pmatrix},
\end{equation}

yielding the 2D coordinates:

\begin{equation}
    \begin{split}
        \hat{x} &= c_{xy}x + c_tt\\
        \hat{y} &= c_{xy}y + c_tt,
    \end{split}
\end{equation}

for scaling constants $c_{xy}$, $c_t$.

\end{itemize}  

\noindent \underline{\emph{Feature embedding}}: After establishing the arrangement of the nodes, the next step is to encode the node features into the quantum system. Here we discuss two possible approaches, where the first one encodes the event features onto the initial state and the second encodes the features using the local detuning term.  

\begin{enumerate}
    \item \emph{Initial state encoding}: In this method, the node features are encoded into the initial state of the atoms $|\psi_s^{\text{init}}\rangle = |x_{s,1}^{\text{init}}\rangle \otimes \cdots  \otimes |x_{s,N_{\text{qubit}}}^{\text{init}}\rangle$. \NOTE{This is achieved by locally driving selected qubits from the ground state $|0\rangle$ to the excited state $|1\rangle$ prior to the quantum graph processing step.} In particular, each atom $i$ corresponding to a new event $e_i$ is initialized as a function of the polarity $p_i$ of the corresponding event $e_i$ as
\begin{equation}\label{eq:new_state}
|x_{s,i}^{\text{init}}\rangle=\begin{cases}
|0\rangle & \text{ if }p_{i}=\NOTE{+1}\\
|1\rangle & \text{ if }p_{i}=\NOTE{-1}.
\end{cases}
\end{equation}
 In contrast, any atom $i$ corresponding to a legacy event is initialized based on the result $x^\mathrm{fin}_{s-1,i} \in \{0,1\}$ of the measurement of the $i$-th atom done at the end of the previous frame. \NOTE{More precisely, as discussed in Sec. \ref{sec:qa-aegnn:readout}, we carry out $N_{\text{shots}}$ measurements, and set 
\begin{equation}\label{eq:prev_state}
|x_{s,i}^{\text{init}}\rangle= |\bar{x}_{s-1,i}^\mathrm{fin}\rangle,
\end{equation}
where $\bar{x}_{s-1,i}^\mathrm{fin} \in \{0,1\}$ is the bit obtained more frequently among the $N_{\text{shots}}$ measurements of atom $i$.

}

 \item \emph{Detuning encoding}: Another possible approach is to encode input information into the local detuning term $\mu_i\delta^{\text{local}}(t)$ from (\ref{eq:detuning}). In \cite{kornjaca2024large}, this approach was used to encode real-valued inputs, whereas in the proposed QA-AEGNN the input corresponds to the node-specific binary polarities $p_i$. \NOTE{With this embedding technique, the qubits are initialized in the ground state $|\psi_s^{\text{init}}\rangle = |0\rangle^{\otimes N_{\text{qubits}}}$, removing the need for additional computation prior to Hamiltonian evolution.} Specifically, setting $\delta^{\text{local}}(t) = \delta^{\text{local}}$, 
 the node feature is embedded in the site modulation coefficient $\mu_i$ using 
 \begin{equation}\label{eq:new_detuning}
 \mu_i = \begin{cases}
     0 & \text{ if } p_i=\NOTE{+1}\\ 
     1 & \text{ if } p_i=\NOTE{-1}
 \end{cases},\end{equation} for new events and
 \begin{equation}\label{eq:prev_detuning}
 \mu_i = \bar{x}_{s-1,i}^{\text{fin}}
 \end{equation} for legacy events. \NOTE{Note that, unlike initial state encoding, detuning encoding requires local driving terms during the quantum time evolution.}
\end{enumerate}

\subsection{Quantum Graph Processing: Hamiltonian Design for Native Analog Graph Neural Computation}

With the atoms positioned to represent the graph as explained in the previous subsection, the system evolves under the Rydberg Hamiltonian for some period of time $\Delta \tau$. The quantum evolution carries out a computation akin to the forward pass of the AEGNN on the input graph. \NOTE{In fact, the Rydberg Hamiltonian's interaction term $\sum_{i<j} V_{ij} n_i n_j$ in \eqref{eq:Rydberg} couples adjacent nodes via the Rydberg blockade phenomenon. This inhibitory effect mimics the additive operation on the AEGNN update rule \eqref{eq:aggregate}. At the same time, the first two terms in \eqref{eq:Rydberg} provide a way to mix or communicate information between states.}

To modulate the quantum evolution, we apply a constant Rabi frequency $\Omega(t) = \Omega$ and a fixed global laser detuning $\delta^{\text{global}}(t) = \delta^{\text{global}}$ over an adjustable time period $\Delta \tau_s$. \NOTE{The constant square wave signals serve as an approximation of the realistic laser pulses employed by the neutral-atom hardware. Nevertheless, the impact of realistic waveforms on the result is expected to be minimal compared to other noise sources \cite{albrecht2023quantum}.} The Rabi frequency $\Omega$ and the global laser detuning $\delta^{\text{global}}$, together with the local detuning $\delta^{\text{local}}$, are used to form the Hamiltonian $H(t) = H$ based on (\ref{eq:Rydberg}). The time duration controls the degree of feature information, playing a role analogous to the  message passing step in a classical GNN. Accordingly, we define the set of tunable parameters $\{\Delta\tau_s\}_{s= 1}^S$, representing the pulse duration in each window $\mathcal{W}_s$.

With these controls, the system evolves under a time-independent Hamiltonian $H$, constructing the final state as follows: 
\begin{equation}\label{eq:final_state}
    |\psi^{\text{fin}}_s\rangle = e^{-iH\Delta\tau_s} |\psi^{\text{init}}_s\rangle.
\end{equation}

We finally note that multiple instances of the same network may be run in parallel on the hardware if $N_{\text{qubit}}$ is small, enabling ensemble predictions or statistical sampling.

\subsection{Node Embedding Readout and Decision}\label{sec:qa-aegnn:readout}

After the programmed analog evolution completes, i.e., at the end of the evolution period $\Delta \tau_s$, \NOTE{the system is measured, in order to extract classical information. We specifically perform measurements in the computational basis on the final state $|\psi_s^{\text{fin}}\rangle$ to obtain the bit $x_{s,i}^\mathrm{fin}\in\{0,1\}$ for each atom $i = 1,\dots,N_{\text{qubit}}$ during time window $\mathcal{W}_s$. State preparation, Hamiltonian evolution, and measurement steps are repeated $N_{\text{shots}}$ times, also known as shots. This yields the empirical probabilities $q_{s,i}^\mathrm{fin}\in[0,1]$ for $i = 1,\dots,N_{\text{qubit}}$. Classical post-processing is then applied to the empirical probabilities, to evaluate the feature vector $z_s$ for the current window using a post-processing function $F(\cdot)$ as  
\begin{equation}\label{eq:postprocess}
    z_s = F(\{q^{\mathrm{fin}}_{s,i}\}_{i=1}^{N_{\text{qubit}}}).
\end{equation}

Note that the features $z_s$ in \eqref{eq:postprocess} consist only of the result in the final layer $L$ due to the state collapse caused by quantum measurement. This differs from AEGNN, in which the features are concatenated over all GNN layers. In future research, it may be possible to extend QA-AEGNN to support weak measurements on ancilla qubits at intermediate steps of the quantum evolution. This would allow the extraction of additional information during Hamiltonian evolution,  more closely mimicking the layered structure of a classical GNN.}
 
\NOTE{Following an approach similar to reference \cite{henry2021quantum}, the function in \eqref{eq:postprocess} is the histogram of the number of atoms measured in the excited state. Accordingly, the vector $z_s$ consists of $N_{\text{qubit}} + 1$ elements, $z_{s}^j$, with $j \in \{0,1,\dots,N_{\text{qubit}}\}$, where $z_{s}^j$ counts the fraction of the $N_{\text{shots}}$ measurements in which $j$ atoms were measured in the excited state, satisfying the condition

\begin{equation}
\sum_{i = 1}^{N_{\text{qubit}}}x_{s,i}^{\text{fin}}  = j.
\end{equation}
}

The resulting features $z_s$ are then fed into a classical machine learning model that depends on the given application. For instance, \cite{schaefer2022aegnn} considers classification and object detection tasks. In this work, we focus on classification; thus $z_s$ serves as an input to a classifier model $f_s(\cdot)$, which produces a decision at the end of each \NOTE{window $\mathcal{W}_s$}\NOTE{, denoted as $\hat{k}_s$}. In order to make a final classification decision \NOTE{$\hat{k}$}, we consider the majority rule (\ref{eq:majority_rule}), as also employed in the classical framework.

\begin{algorithm}
    \caption{QA-AEGNN}
    \label{algo:QAEGNN}
    \textbf{Input}: Event sequence $E = (e_1, e_2, \dots, e_N)$, with event $e_i = (x_i, y_i,t_i, p_i)$ where $t_1 \leq t_2 \leq \dots \leq t_N$ and polarity $p_i \in \{-1, +1\}$; inter-atomic distance $R > 0$; time window duration $\Delta T$; window shift $\tau$ where $\Delta T = m\tau$ for some integer $m$; set of parameters $\{\Delta\tau_s\}_{s= 1}^S$; Rabi constant $\Omega$; global laser detuning $\delta^{\text{global}}$feature embedding type $\text{feat} = \{\text{init\_state},\text{local\_detun}\}$; number of layers $L$; local detuning $\delta^{\text{local}}$; device dimension $D$; if 3D device, scaling factors $c$ and $\beta$; if 2D device, matrix $M$; number of measurement shots $N_{\text{shots}} \geq 1$.
    \begin{algorithmic}[1]
        \State $E_{0} \leftarrow \emptyset$ \Comment{Initialize event set}
            \For {$s = 1, 2, \dots, S$}
            \State $\mathcal{W}_s \leftarrow [s\tau - \Delta T, s\tau]$ \Comment{Form time window}
            \State $E_s \leftarrow \{ e_i : e_i = (x_i, y_i, t_i, p_i)\in E, t_i \in \mathcal{W}_s\}$ \Comment{Form event set}
            \If{$D = 3$}
                \State $\{r_i\}_{e_i \in E_s} \leftarrow \{ c \cdot(x_i, y_i, \beta t_i): e_i = (x_i, y_i, t_i, p_i) \in E_s\}$ \Comment{3D device}
            \Else
                 \State $\{r_i\}_{e_i \in E_s} \leftarrow \{ M(x_i, y_i, t_i)^T: e_i = (x_i, y_i, t_i, p_i) \in E_s\}$ \Comment{2D device}
            \EndIf
            \MRepeat {$N_{\text{shots}}$ times}
            \If{$\text{feat} = \text{init\_state}$} \Comment{Initial state embedding}
                \State Initialize $\{|x_{s,i}^{\text{init}}\rangle\}_{i =1}^{N_{\text{qubit}}}$ using equations (\ref{eq:new_state})-(\ref{eq:prev_state})
                \State $|\psi_{s}^{\text{init}}\rangle \leftarrow |x_{s,1}^{\text{init}}\rangle\otimes\dots\otimes|x_{s,N_{\text{qubit}}}^{\text{init}}\rangle$
                \State Set $\{\mu_i \leftarrow 0\}_{i = 1}^{N_{\text{qubit}}}$
            \Else \Comment{Local detuning embedding}
                \State $|\psi_s^{\text{init}}\rangle \leftarrow |0\rangle^{\otimes N_{\text{qubits}}}$
                \State Initialize $\{\mu_i\}_{i=1}^{N_{\text{qubit}}}$ using equations (\ref{eq:new_detuning})-(\ref{eq:prev_detuning})
            \EndIf
            \State $H = \sum_{i=1}^{N_{\text{qubit}}} \left [\frac{\Omega}{2} \left( |1_i\rangle \langle 0_i|  + |0_i\rangle \langle 1_i| \right) 
\; - \; (\delta^{\text{global}} + \delta^{\text{local}}\mu_i) n_i \right ]
\; + \; \sum_{i < j} \frac{C_6}{\| {r}_i - {r}_j \|^6} n_i n_j$ \Comment{Form $H$}
            \State $|\psi_s^{\text{fin}}\rangle \leftarrow e^{-iH\Delta\tau_s}|\psi_s^{\text{init}}\rangle$ \Comment{Apply the quantum evolution} 
        \State Form $\{x^{\mathrm{fin}}_{s,i}\}_{i=1}^{N_{\text{qubit}}}$ from measurement of state $|\psi_s^{\text{fin}}\rangle$ \Comment{Measure the system}
        \EndRepeat
        \State \NOTE{Evaluate the empirical probability $\{q_{s,i}\}_{i=1}^{N_{\text{qubit}}}$ using the $N_{\text{shots}}$ measurements $ \{x_{s,i}^{\text{fin}}\}_{i = 1}^{N_{\text{qubit}}}$}
        \State $z_s \leftarrow F\left(\{q^{\mathrm{fin}}_{s,i}\}_{i=1}^{N_{\text{qubit}}}\right)$ \Comment{Construct the histogram of excited atoms}
        \State \NOTE{$\hat{k}_s \leftarrow  f_s(z_s)$ \Comment{Class prediction at \NOTE{window $\mathcal{W}_s$}}}
        \EndFor
    \end{algorithmic}
     \hspace*{\algorithmicindent}\textbf{Return:} \NOTE{Classification decision $\hat{k} = \arg \max_{\kappa} \sum_{s = 1}^S \mathbbm{1}(\hat{k}_s = \kappa)$}
\end{algorithm}

\subsection{Training QA-AEGNN}\label{sec:training}

For each \NOTE{window $\mathcal{W}_s$}, as summarized in Algorithm \ref{algo:QAEGNN}, the operation of QA-AEGNN depends on  the parameters $(\tau_s, \theta_s)$, where $\tau_s$ is the tunable pulse duration and $\theta_s$ represents the learnable parameters of the classifier model $f_s(\cdot)$. As for AEGNN (see Sec. \ref{sec:aegnn_training}), we optimize these parameters by using a training dataset $\mathcal{D}$ comprised of event-label pair $(E_s, k)$, where $E_s$ is the sequence of events (\ref{eq:eventwins}) for \NOTE{window $\mathcal{W}_s$}  and $k$ is a classification label. 

As outlined in Algorithm \ref{algo:QAEGNN}, in every \NOTE{window $\mathcal{W}_s$}, the feature vector $z_s$ depends on the input $E_s$ and the time-duration parameter $\tau_s$. To make this dependence explicit, henceforth we use the notation $z_s(E_s | \tau_s)$. Subsequently, the classifier takes as input the features $z_s(E_s | \tau_s)$, and uses the classifier parameters $\theta_s$ to produce the classification  prediction  $f_s(z_s(E_s|\tau_s) | \theta_s)$. 

We propose to employ a quantum-classical hybrid training loop 
to jointly train \NOTE{$\tau_s$ and $\theta_s$. Accordingly, training is performed using classical machine learning methods, while we use quantum resources to implement inference.} Specifically, using a classification loss function $\ell(\cdot,\cdot)$, the parameters $(\tau_s, \theta_s)$ are optimized by addressing the empirical risk minimization problem
\begin{equation}\label{eq:loss_qaegnn}
    \min_{\tau_s, \theta_s} \mathbb{E}_\mathcal{D}[\ell(k, f_s(z_s(E_s|\tau_s) | \theta_s))].
\end{equation}
In practice, this problem is solved by using a grid of candidate values for the quantum evolution $\tau_s$,  addressing problem (\ref{eq:loss_qaegnn}) separately over the classifier parameter $\theta_s$ for each fixed value $\tau_s$.

\NOTE{
\subsection{Real-time Quantum Inference}

We conclude this section by examining conditions under which real-time inference on a quantum device could become feasible, by selecting an appropriate value for the time window duration $\Delta T$.

Two complementary constraints govern the admissible range of $\Delta T$. First, the number of events processed within a single time frame $\Delta T$ is bounded above by the number of available atoms on the quantum device. Formally, letting $R_{\text{event}}$ denote the event rate (event/s), and $N_{\text{atom}}$ the number of atoms in the quantum register, this capacity constraint reads $N_{\text{atom}} \geq R_{\text{event}} \cdot \Delta T$, which yields the following upper bound:
\begin{equation}\label{eq:delta_t_upper}
\Delta T \leq \frac{N_{\text{atom}}}{R_{\text{event}}}.
\end{equation}

Second, all quantum operations associated with a given window $\mathcal{W}_s$ must complete before the subsequent window $\mathcal{W}_{s+1}$ begins processing. Denoting by $T_{\text{SPAM}}$ the time required for state preparation and final-state measurement, and recalling that $N_{\text{shots}}$ circuit executions are performed per window, the window shift $\Delta T/2$ — which governs the temporal separation between consecutive frames — must satisfy the lower bound
\begin{equation}\label{eq:delta_t_lower}
\frac{\Delta T}{2} \geq \left(T_{\text{SPAM}} + \Delta\tau\right) N_{\text{shots}}.
\end{equation}
Substituting the lower bound on $\Delta T$ from (\ref{eq:delta_t_lower}) into (\ref{eq:delta_t_upper}), one obtains the following necessary condition on the event rate $R_{\text{event}}$ for real-time inference to be achievable
\begin{equation}
R_{\text{event}} \leq \frac{N_{\text{atom}}}{2\left(T_{\text{SPAM}} + \Delta\tau\right) N_{\text{shots}}}.
\end{equation}
Applications whose event rates satisfy this bound are therefore suitable candidates for real-time quantum inference. For example, assuming a platform with $N_{\text{atom}} = 256$ atoms \cite{wurtz2023aquila}, $N_{\text{shots}}=10$ shots and $T_{\text{SPAM}} + \Delta\tau = 0.1\ \text{s}$ \cite{wurtz2023aquila}, the rate could be as high as $R_{\text{event}} = 128 \ \text{event/s}$. Larger event rates may be attained as neutral-atom platforms scale up in terms of number of atoms and shot rate $1/(T_{\text{SPAM}} + \Delta\tau)$.
}

\section{Experiments}\label{sec:experiments}

In this section, we compare the proposed QA-AEGNN with the classical AEGNN \cite{schaefer2022aegnn} on two classification benchmarks. The first is the synthetic graph classification task introduced in \cite{albrecht2023quantum}, while the second is based on a standard neuromorphic dataset \cite{orchard2015converting}.

\subsection{Tasks}

\subsubsection{Graph Classification}

Following \cite{albrecht2023quantum}, we address the problem of distinguishing graphs based on the sequential observation of node coordinates. We also consider an augmented dataset encompassing also a binary per-node anomaly flag, serving as the event polarity. 

\begin{figure}
    \centering
    \includegraphics[width=0.7\textwidth]{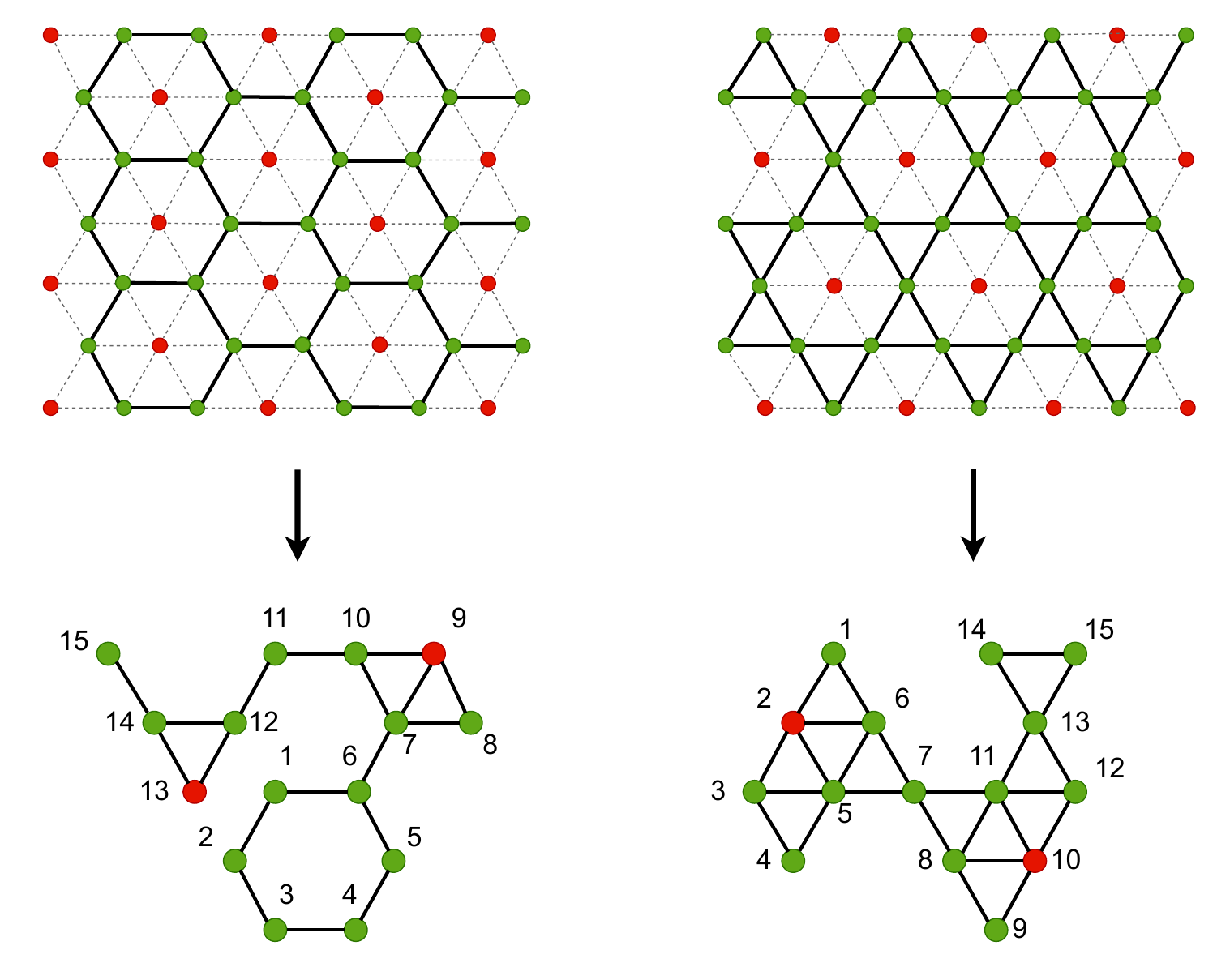}
    \caption{Illustration of the graph classification task \cite{albrecht2023quantum}: Graphs in each class are generated as shown in the top panel. In the first class (left), nodes of the graph are generated by following a honeycomb-like path with probability, proportional to weight $w_0 = 1$ (green nodes), while allowing deviations (red nodes) with a probability proportional to the weight $w < 1$. In the second class (right), graphs are generated by following a kagome-like path using the same probabilities. Each event $e_i$ has spatio-temporal coordinates $X_i = (x_i, y_i, t_i)$, where $(x_i, y_i)$ denote the node position in the 2D lattice on a unit lattice and time $t_i$ indicates the integer visitation order. All events have polarity $p_i = 1$.}
    \label{fig:synthetic}
\end{figure}

As illustrated in the top panel of Fig. \ref{fig:synthetic}, the graphs are constructed as subgraphs of a 2D triangular lattice grid with edges of normalized length equal to 1. The subgraphs are built through random walks on the nodes of the lattice, with distinct preferential transitions for the two classes. Specifically, one class of graphs is generated by preferentially selecting the next node along a honeycomb pattern, as shown in the top-left panel of Fig. \ref{fig:synthetic} (solid  lines), while allowing possible deviations from the template pattern (dashed lines). As illustrated in the top right panel of Fig. \ref{fig:synthetic}, the other class is constructed in the same way, but favoring a path that traces a kagome-like structure (solid lines). 

In more detail, each graph in either class is generated by initiating a random walk from a randomly chosen point on the per-class sublattice (green vertices in the top panels of Fig. \ref{fig:synthetic}), which is assigned coordinates $(x_0 = 0, y_0 = 0)$. The next node $(x_1, y_1)$ is selected to follow the class template, i.e., to be an adjacent green node, with a probability proportional to weight $w_0 = 1$, while deviations from the template, corresponding to red nodes, are allowed with a lower probability proportional to weight $w < 1$. The process is repeated for all coordinates $(x_i, y_i)$, until $N$ unique sites are visited, forming an event graph of size $N$. 

By this construction, the parameter $w$ determines how closely the random walk follows the per-class template. In fact, for a node with $N_g$ neighboring nodes within the per-class pattern (green nodes), and $N_r$ neighbors outside the per-class pattern (red nodes), the probability of transitioning to a \NOTE{in-template node} is given by $N_g/(N_g + wN_r)$, while the probability of reaching \NOTE{out-of-template node} is $wN_r/(N_g + wN_r)$. This way, with a zero weight $w = 0$, the graphs strictly adhere to the class template, forming easily separable classes. Increasing the value of the weight $w$ leads to a higher chance to \NOTE{explore nodes that do not follow the graph class template}, making classification progressively more challenging.

We consider two data structures based on this construction: \begin{itemize} \item \emph{Coordinate-only data:}  The spatio-temporal location $X_i = (x_i, y_i, t_i)$ of an event $e_i$ is formed by the node coordinates $(x_i, y_i)$, and by an integer time index $t_i$ describing the order in which the site with coordinates $(x_i, y_i)$ is first visited by the random walk generating the graph. Polarity is assumed to be $p_i = 1$ for every event $e_i$. 
\item \emph{Polarity-augmented data:} Each event $e_i=(x_i, y_i, t_i,p_i)$ includes not only the spatio-temporal location $X_i = (x_i, y_i, t_i)$ but also a binary flag $p_i$ indicating whether the corresponding graph node follows the main pattern (green nodes in Fig. \ref{fig:synthetic}) or not (red nodes).
\end{itemize}

\subsubsection{Neuromorphic Data Classification}

We also consider a binary classification task on a neuromorphic dataset, namely N-MNIST \cite{orchard2015converting}. The N-MNIST dataset extends the standard MNIST dataset by converting the static images of handwritten digits into spatio-temporal event-based representations. Transformation is achieved by recording the images using an asynchronous time-based image sensor  mounted on a pan-tilt camera platform, capturing the MNIST images for $320\  \text{ms}$ while performing controlled sensor movements. Events $e_i$ are directly encoded as an output of the sensor, where the spatial coordinates $(x_i,y_i)$  correspond to pixel positions, $t_i$ denotes the timestamp associated with each event, and the polarity $p_i$ indicates the sign of the intensity change of a pixel. As for the previous setting, we consider datasets that exclude or include the polarity $p_i$.

Considering the task of binary classification, we focus on distinguishing between the digits $0$ and $1$ in the dataset. Rather than using the entire image, the experiment considers only the central region of each \NOTE{$34 \times 34$} image, specifically events whose $x_i$ and $y_i$ coordinates lie within the range $[10, 20)$. The data from each image is further subsampled by selecting only $5\%$ of the events within the specified region.

\subsection{Implementation}

As the classifier model $f_s$ we adopt a support vector machine (SVM) with a kernel derived from the Jensen-Shannon ($\text{JS}$) divergence, \NOTE{as in \cite{henry2021quantum}.} Accordingly, given histograms $z_s$ and $z'_s$, corresponding to the extracted features from graphs different graphs, the kernel is defined as $\kappa (z_s,z_s') = \exp(-\text{JS}(z_s, z_s'))$,  where 
\begin{equation}
\text{JS}(z_s, z_s') = \frac{1}{2}\text{KL}\left (z \middle\| \frac{z + z'}{2} \right) + \frac{1}{2}\text{KL}\left (z' \middle\| \frac{z + z'}{2} \right)
\end{equation}
is the JS divergence and $\text{KL}(z_s||z'_s) = \sum_i z_{s,i}\log(z_{s,i}/z'_{s,i})$ is the Kullback-Leibler (KL) divergence. 
To address the empirical risk minimization problems  in (\ref{eq:loss}) and (\ref{eq:loss_qaegnn}), we choose the model classification error as a loss function $\ell(\cdot,\cdot)= 
     \mathbbm{1}( k \neq f_s(z_s(E_s|\alpha_s) | \theta_s)), $
where $\mathbbm{1}(\cdot)$ is the indicator function.

Moreover, for our first experiment, we use a window size of $\Delta T = 10$ and an inter-window shift of $\tau = 5$, while for our second experiment we use $\Delta T = 30 \ \mu\text{s}$ and $\tau = 15 \ \mu\text{s}$.

For AEGNN, we apply the message passing scheme for $L = 4$ layers, and in each layer $l$ the features are quantized into $B = 10$ bins using the intervals $(a^l, b^l) = (\min_{i \in \mathcal{V}_s}(h_{s,i}^l),  \max_{i \in \mathcal{V}_s}(h_{s,i}^l))$ for $l=1,\dots, L$. The parameter $\alpha_s$, used to control the slope of the sigmoid function $\sigma$, is jointly optimized with the classifier by selecting the value that maximizes the accuracy over 20 discrete candidates uniformly distributed in the range $\NOTE{\alpha_s} \in [0.1, 2]$. For the \NOTE{graph} dataset, edges in the AEGNN are formed within a radius $R = 1$, while for the neuromorphic dataset, the radius is set to $R = 5$. Time rescaling is applied with $\beta = 0$ for the \NOTE{graph} task, and $\beta = 1$ for the neuromorphic MNIST dataset. 

\NOTE{For QA-AEGNN, we set the Rabi frequency $\Omega/2\pi = 1$ MHz and the laser detuning $ \delta/2\pi = 0.7$ MHz.} The interaction coefficient in (\ref{eq:blockade}) is set to $C_6 =  865723.02$. This yields the Rydberg blockade radius $R_b = \left ( C_6/\hbar \Omega \right ) ^{\frac{1}{6}} \approx 7.19\  \mu \text{m}$. \NOTE{The quantum evolution is performed $N_\text{shots} = 100$ times.} Events are embedded on a 2D array of atoms using the linear transformation from (\ref{eq:linear_transformation}). For the \NOTE{graph} task, we set $c_{xy} = R_b/2\sqrt{2}$, ensuring that neighboring nodes lie within the Rydberg blockade radius, as described in \cite{albrecht2023quantum}, and set $c_t = 0$, matching the AEGNN setting $\beta = 0$. For the NMNIST dataset, we fix $c_{xy} = R_b/\sqrt{2}$, to facilitate a dominant van der Waals interaction between pixels within a $5\times5$ region around each pixel, and set $c_t = 1$, matching $\beta = 1$ in the classical setting. The evolution time $\Delta\tau_s$ for every \NOTE{window $\mathcal{W}_s$} is optimized jointly with the SVM classifier to minimize the loss as discussed in Sec. \ref{sec:training}. 

The simulation of the quantum evolution is done using the Pulser python library \cite{silverio2022pulser}, together with the state-vector emulator, emu-sv, for systems that contain less than $25$ qubits and Matrix Product State tensor based emulator, emu-mps, for larger arrays \cite{bidzhiev2025efficient}. For both tasks, results are reported in terms of the median classification accuracy, as well as the 25th and the 75th percentile which are evaluated across multiple trial runs. 

 In addition to the noiseless simulation of the quantum evolution, we consider two different noise models. The first noise model accounts only for state preparation and measurement (SPAM) errors, which are independent errors across all qubits \cite{yu2025efficient}, modeled within the classical readout. The false positive error rate $\epsilon_p$, corresponding to falsely measuring $1$ when the true outcome is $0$, is set to $\epsilon_p = 0.025$, and the false negative error rate $\epsilon_n$, corresponding to measuring $0$ instead of $1$, is $ \epsilon_n = 0.1$. The second noise model includes not only SPAM errors but also quantum noise. Following \cite{bravo2022quantum}, the decoherence of the quantum system, characterizing the decay to the ground state, is modeled using a relaxation time $T_1$, while the loss of phase coherence between the basis states during the quantum evolution is modeled by dephasing time $T_2$. The relaxation rate and the dephasing rate are set to $1/T_1 = 0.01 \ \mu\text{s}^{-1}$ and $1/T_2 = 0.22 \ \mu\text{s}^{-1}$. 

Using the proposed noise models, we consider two different settings for training:

\begin{enumerate}
    \item \emph{Noise-agnostic training}: Training is done assuming a noiseless system, and noise is applied only during inference;
    \item \emph{Noise-aware training}: Training is done on a system that is affected by the same noise model to be encountered at test time.
\end{enumerate}

\subsection{Results for the Graph Classification Task}

\subsubsection{Feature encoding}

To start, we demonstrate the training process for QA-AEGNN by focusing on the impact of the quantum evolution time $\tau_s$ for a given window index $s$, here $s = 1$. We execute $50$ trials on different balanced datasets, consisting of $200$ event graphs, where each graph contains $N = 100$ events. We use $80\%$ of the graphs for training, and the remaining $20\%$ graphs are used for evaluation. The weights of the SVM classifier are trained to minimize the training loss as discussed above.

 \begin{figure}
    \centering
    \begin{subfigure}{.49\textwidth}
        \centering
        \includegraphics[width=\linewidth]{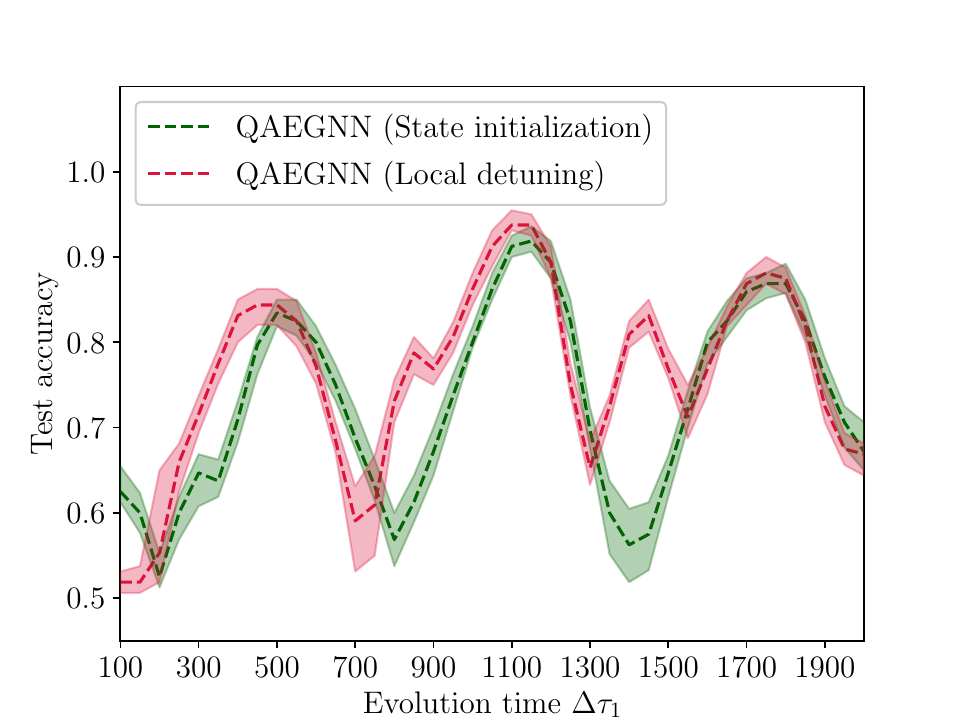}
        \label{fig:q_evolution_detuning_a}
    \end{subfigure}
    \begin{subfigure}{.49\textwidth}
        \centering
        \includegraphics[width=\linewidth]{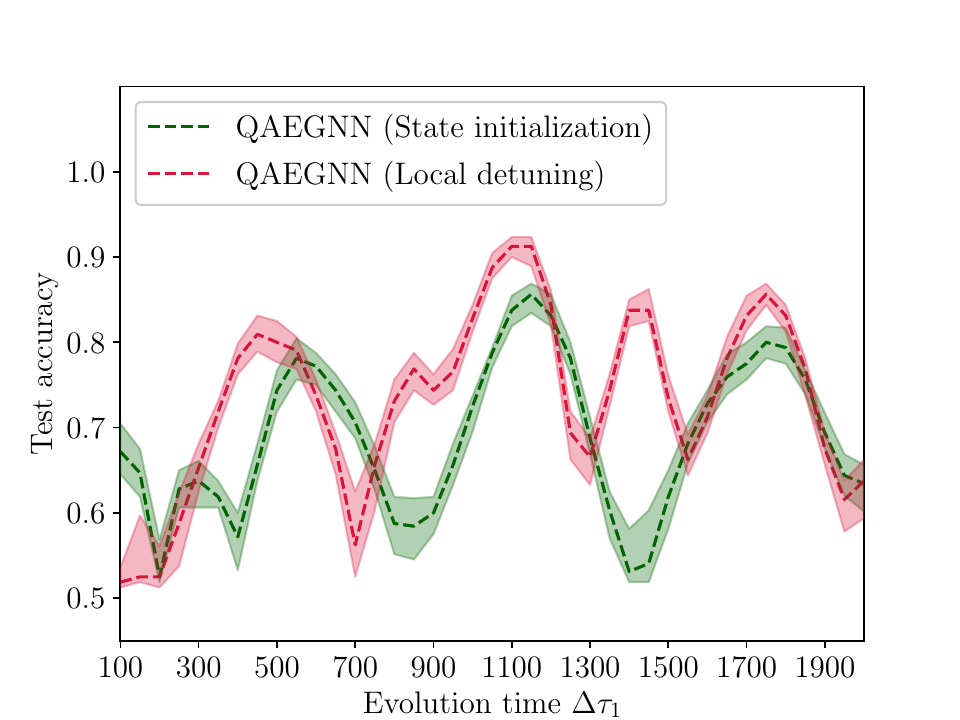}
        \label{fig:q_evolution_detuning_b}
    \end{subfigure}
    \caption{Graph classification task: Training accuracy of QA-AEGNN on the polarity augmented dataset for state initialization (green) and local detuning (red) feature encoding schemes as a function of the quantum time evolution $\Delta \tau_s$ for the initial window $s = 0$ for $w = 0.1$ (left) and $w = 0.2$ (right).}
    \label{fig:q_evolution_detuning}
\end{figure}

To this end, we plot in Fig. \ref{fig:q_evolution_detuning} the training accuracy as a function of  the quantum time evolution using the polarity-augmented dataset for two distinct values of the similarity weight parameter, namely $w = 0.1$, representing easily distinguishable graphs, and $w = 0.2$, corresponding to  a more challenging classification problem.

As shown in the figure, the accuracy fluctuates as the evolution time $\Delta \tau_s$ increases, with both encoding methods obtaining optimal performance at approximately $1.1 \ \mu \text{s}$. When the task of distinguishing the graphs is simpler ($w = 0.1$), the two encoding strategies achieve nearly the same peak accuracy, with optimal accuracies being $0.919$ and $0.938$ for state initialization and the local detuning, respectively. However, for the more challenging task with $w = 0.2$, the local detuning feature embedding outperforms the state initialization encoding, achieving a median optimal training accuracies of $0.913$ and $0.856$, respectively. This suggests that retaining polarity information throughout the quantum evolution via the detuning term may enhance the performance of the quantum model.

  \begin{figure}
    \centering
    \begin{subfigure}{.49\textwidth}
        \centering
        \includegraphics[width=\linewidth]{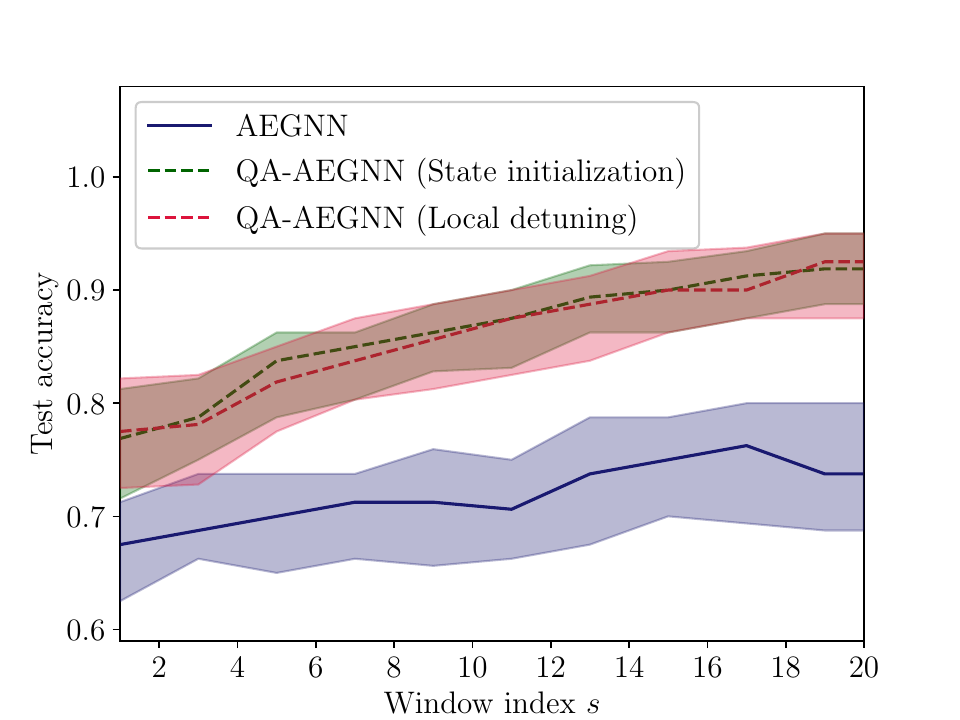}
        \label{fig:detuning_w_0.1_a}
    \end{subfigure}
    \begin{subfigure}{.49\textwidth}
        \centering
        \includegraphics[width=\linewidth]{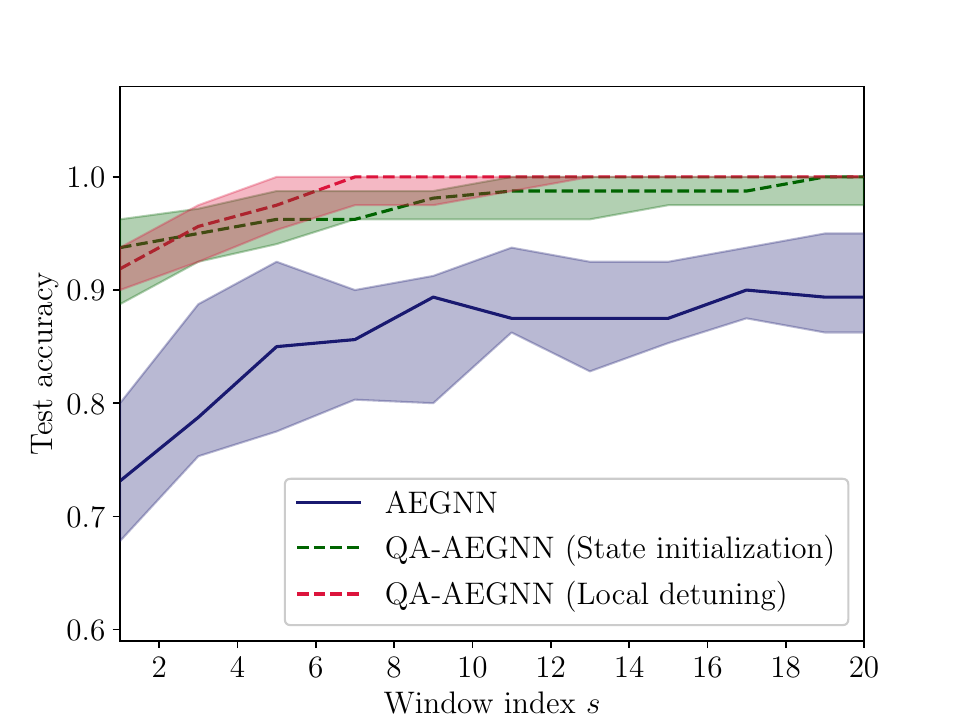}
        \label{fig:detuning_w_0.1_b}
    \end{subfigure}
    \vfill
    \begin{subfigure}{.49\textwidth}
        \centering
        \includegraphics[width=\linewidth]{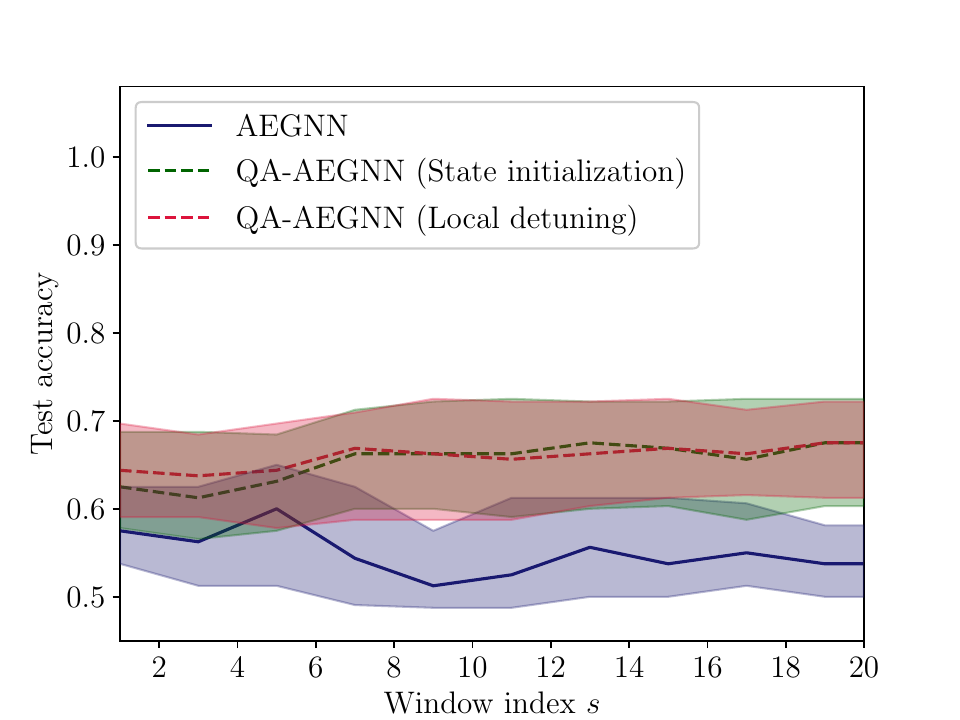}
        \label{fig:detuning_w_0.2_a}
    \end{subfigure}
    \begin{subfigure}{.49\textwidth}
        \centering
        \includegraphics[width=\linewidth]{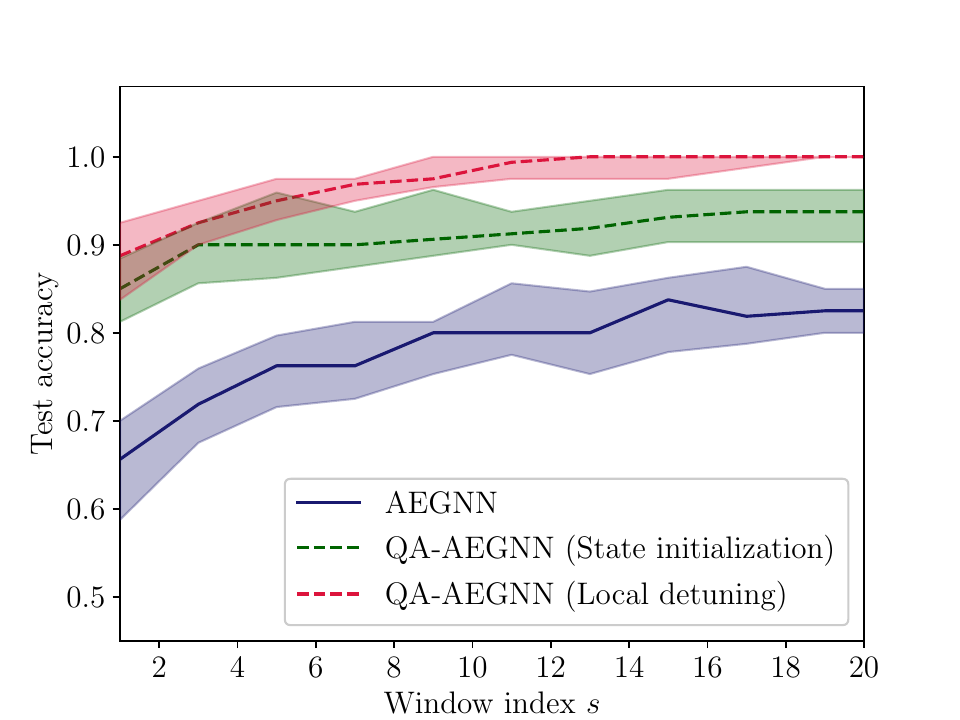}
        \label{fig:detuning_w_0.2_b}
    \end{subfigure}
    \caption{Graph classification task: Test accuracy of AEGNN (blue), QA-AEGNN with quantum state initialization encoding (green), and QA-AEGNN with local detuning encoding (red) as a function of window index $s$ for $w = 0.1$  on coordinate-only data (top left) and polarity-augmented data (top right), and for $w = 0.2$ on coordinate-only data (bottom left) and polarity-augmented data (bottom right).}
    \label{fig:q_detuning}
\end{figure}

\begin{table}
    \centering
    \begin{tabular}{|c||p{2.7cm}| p{2.7cm}| p{2.7cm}| p{2.7cm}|}
        \hline
         \NOTE{\backslashbox{Node embedding}{Graph data}} & \vspace{-0.64cm}\NOTE{Coordinate-only; $w= 0.1$}  & \vspace{-0.64cm}\NOTE{Polarity-augmented; $w=0.1$} & \vspace{-0.64cm}\NOTE{Coordinate-only; $w= 0.2$}  & \vspace{-0.64cm}\NOTE{Polarity-augmented; $w=0.2$} \\\hline\hline\
         \NOTE{AEGNN} & 0.725 (0.675 - 0.8) & 0.888 (0.85 - 0.95) & 0.525 (0.5 - 0.6) & 0.825 (0.8 - 0.85) \\
         \NOTE{QA-AEGNN (State Initialization)} & 0.913 (0.875 - 0.95) & 1.0 (0.975 - 1.0) & 0.675 (0.575 - 0.725) & 0.95 (0.9 - 0.975) \\
         \NOTE{QA-AEGNN (Local Detuning)}& 0.925 (0.875 - 0.95) & 1.0 (1.0 - 1.0)& 0.675 (0.6 - 0.725) & 1.0 (1.0 - 1.0)\\\hline
    \end{tabular}
    \caption{\NOTE{Graph classification task: Median test accuracy along with the corresponding $25$-th and $75$-th percentile intervals of QA-AEGNN with quantum state initialization encoding and local detuning encoding on the different graph datasets. }\label{table:q_detuning} }
\end{table}

We now turn to comparing the test accuracy of QA-AEGNN and of its classical counterpart, AEGNN, as a function of the window frame index $s$ in Fig. \ref{fig:q_detuning}. The top two panels show results for the simpler classification task ($w = 0.1$), while the bottom panels correspond to the more challenging case ($w = 0.2$). In each case, AEGNN and the two QA-AEGNN embedding schemes are evaluated on both coordinate-only (left panel) and polarity-augmented (right panel) datasets. \NOTE{Additionally, Table \ref{table:q_detuning} presents the median test accuracy of the considered models after the final time frame is considered.}  

As observed in Fig. \ref{fig:q_detuning}, QA-AEGNN consistently achieves higher test accuracy compared to AEGNN under all the investigated cases. In particular, when only the coordinate information is considered, the performance of AEGNN plateaus as more window frames are processed. In contrast, QA-AEGNN continues to improve as more events are considered when $w = 0.1$ (top left panel), while maintaining a higher accuracy in case when $w = 0.2$ (lower left panel). When the polarity-augmented datasets are considered (right panels), the quantum algorithm demonstrates a clear advantage in distinguishing between the two classes, with QA-AEGNN reaching perfect test accuracy using any of the two proposed encoding methods for the simpler task. 

Examining the impact of the feature embedding, we observe that, when only coordinate data is available, both quantum embeddings yield similar accuracy. This is expected since the polarity information is discarded in this case. However, with polarity-augmented data, the local detuning embedding tends to outperform the state initialization encoding, suggesting that preserving the feature information throughout the quantum time evolution improves the generalization capabilities of the model. Finally, the disparity between the different encoding methods is more pronounced when the task of distinguishing between the two classes is more challenging, as shown in the lower right panel of Fig. \ref{fig:q_detuning}.

\subsubsection{Noise Sensitivity}

 \begin{figure}
    \centering
    \begin{subfigure}{.49\textwidth}
        \centering
        \includegraphics[width=\linewidth]{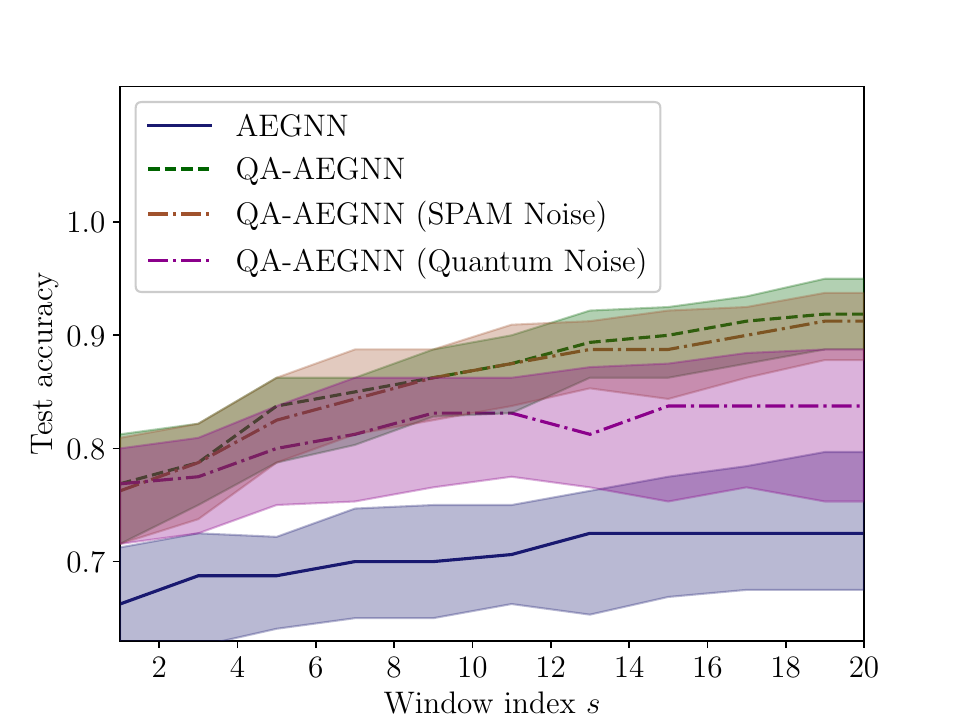}
        \label{fig:synthetic_noise_agnostic}
    \end{subfigure}
    \begin{subfigure}{.49\textwidth}
        \includegraphics[width=\textwidth]{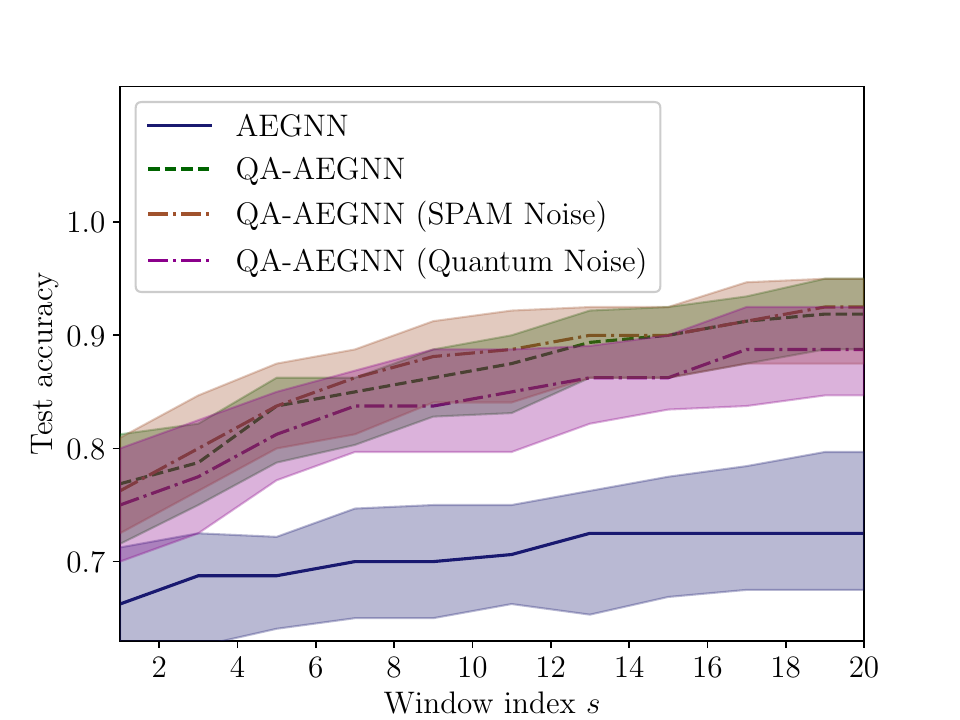}
        \label{fig:synthetic_noise_aware}
    \end{subfigure}
    \caption{Graph classification task: Test accuracy of AEGNN (blue), QA-AEGNN without noise (green), QA-AEGNN with SPAM noise (orange), and QA-AEGNN with quantum noise (purple) as a function of window index $s$ for $w = 0.1$  on coordinate-only data using noise-agnostic training (left) and noise-aware training  (right).}
    \label{fig:synthetic_noise}
\end{figure}

\begin{table}
    \centering
    \begin{tabular}{|c||c | c|}
        \hline
         \NOTE{\backslashbox{Noise model}{Training type}} & \NOTE{Noise agnostic} & \NOTE{Noise aware}  \\\hline\hline\
         \NOTE{AEGNN} & \multicolumn{2}{c|}{0.725 (0.675 - 0.8)}   \\\hline
         \NOTE{QA-AEGNN (Noiseless)} & \multicolumn{2}{c|}{0.9125 (0.875 - 0.95)}   \\\hline
         \NOTE{QA-AEGNN (SPAM noise)} & 0.9 (0.85 - 0.925) &  0.925 (0.875 - 0.95) \\\hline
         \NOTE{QA-AEGNN (Quantum noise)} & 0.85 (0.775 - 0.9)  & 0.9 (0.875 - 0.925) \\\hline
    \end{tabular}
    \caption{\NOTE{Graph classification task: Median test accuracy, along with the corresponding $25$-th and $75$-th percentile intervals of AEGNN, QA-AEGNN without noise, QA-AEGNN with SPAM noise and QA-AEGNN with quantum noise over the noise-agnostic and noise-aware training.} }
    \label{table:q_noise}
\end{table}

We now consider the effect of quantum noise on the accuracy of the quantum model by comparing the performance of QA-AEGNN  and AEGNN, where QA-AEGNN is subject to SPAM and possibly also quantum noise. The resulting test accuracy is shown in Fig. \ref{fig:synthetic_noise} against the window frame index $s$, where the left panel shows the results for  noise-agnostic training and  the right panel considers noise-aware training. In both cases, we consider the coordinate-only classification task with $w = 0.1$. \NOTE{Table \ref{table:q_noise} presents the test accuracy of the considered models after observing all time frames.}

As shown in Fig. \ref{fig:synthetic_noise}, QA-AEGNN maintains higher test accuracy compared to AEGNN across all noise conditions. When only SPAM noise is present, the accuracy of QA-AEGNN closely matches its noiseless counterpart in both training settings. \NOTE{Noise robustness can be partly attributed to the majority rule strategy employed by QA-AEGNN when combining the information from all time frames to predict the class label.} Under additional relaxation and decoherence noise, accuracy decreases slightly relative to the noiseless and SPAM-only cases. Yet, QA-AEGNN still outperforms AEGNN, indicating an advantage even in the presence of realistic noise conditions. With noise-agnostic training, the accuracy gap between QA-AEGNN with quantum noise and the other two configurations widens with increasing window index $s$. In contrast, with  noise-aware training, the gap remains roughly stable across most windows, suggesting that training and testing on the same hardware platform may lead to more consistent performance.

\subsection{Results for the Neuromorphic MNIST Dataset}

  \begin{figure}
    \centering
    \begin{subfigure}{.49\textwidth}
        \centering
        \includegraphics[width=\linewidth]{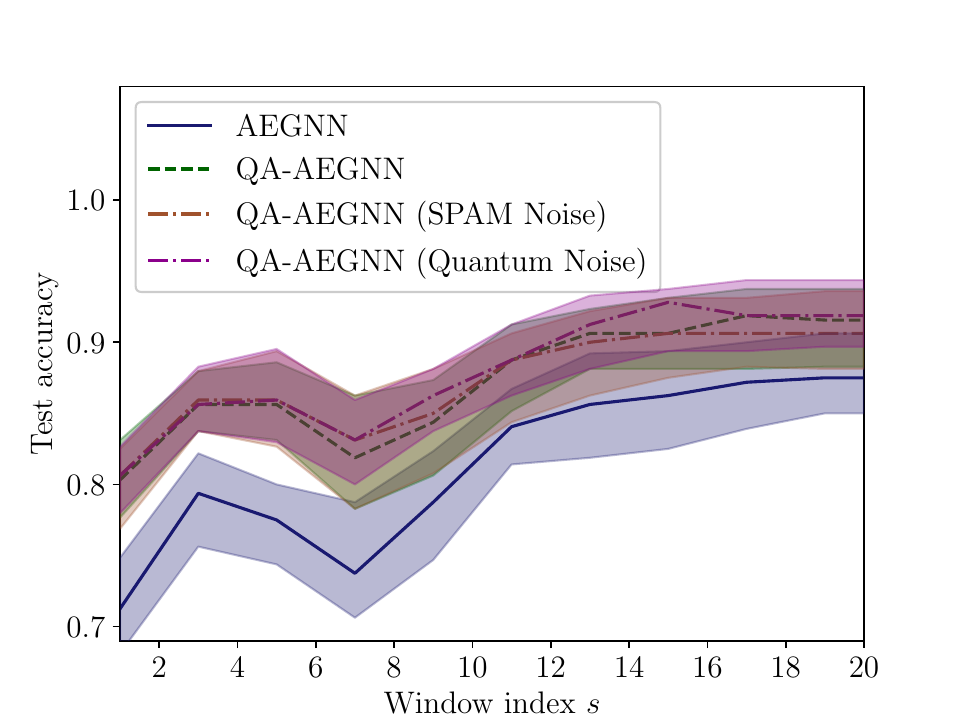}
        \label{fig:neuromorphic_evolution}
    \end{subfigure}
    \vfill
    \begin{subfigure}{.49\textwidth}
        \centering
        \includegraphics[width=\linewidth]{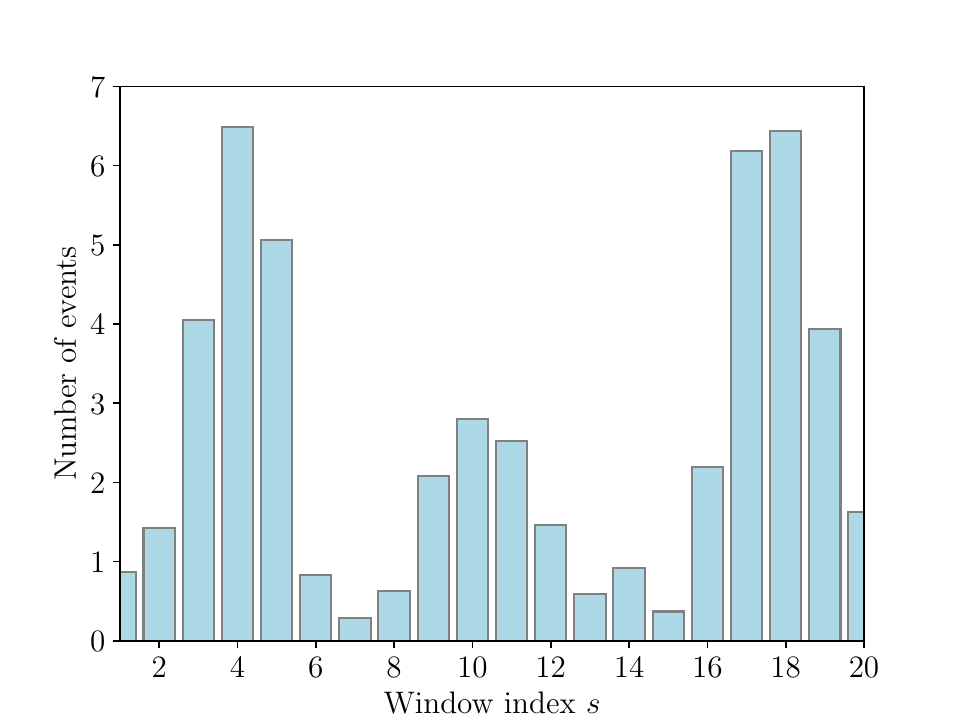}
        \label{fig:neuromorphic_event_count}
    \end{subfigure}
    \caption{NMNIST task: Training accuracy of AEGNN (blue), QA-AEGNN without noise (green), QA-AEGNN with SPAM noise (orange), and QA-AEGNN with quantum noise (purple) as a function of window index $s$ (upper panel) using noise-aware training model. Average number of events in a window $\mathcal{W}_s$, as a function of window index $s$ (lower panel).}
    \label{fig:nmnist_plots}
\end{figure}

We now evaluate QA-AEGNN and AEGNN on the N-MNIST dataset. We execute $50$ trials on differently sampled datasets, each consisting of $400$ event graphs.  As in the previous example, $80\%$ of the samples are used to train the SVM classifier and optimize the set of parameters, and the remaining $20\%$ are used for evaluation.

The upper panel of Fig. \ref{fig:nmnist_plots}  plots the test accuracy of QA-AEGNN  and AEGNN as a function of the window size $s$. For QA-AEGNN, we show the performance in the absence of noise, with SPAM noise, as well as  with SPAM and quantum noise assuming noise-aware training. The lower panel of the figure shows the average number of events per frame, calculated across all neuromorphic images in the datasets. Note that, unlike the previous \NOTE{graph} example, here the number of events can vary significantly from window to window.

As shown in the figure, the test accuracy of QA-AEGNN is higher compared to  AEGNN. Furthermore, consistent with the \NOTE{graph} dataset results, the SPAM-noise and the noiseless QA-AEGNN models do not affect the accuracy of QA-AEGNN in a significant way. An interesting observation is that a notable decline in accuracy is observed for all models between window indices $s=6$ and $s=8$. As seen in Fig. \ref{fig:nmnist_plots}, the average event count  is much lower for index $s$ in this range, indicating  that the performance degradation can be attributed to limited information carried by events in these windows. This decline coincides with the change of direction of the pan-tilt camera platform, used during image recording \cite{orchard2015converting}. A similar lower event count around frame $s=15$ is observed not to have the same impact. This is due to the fact that, by this window index, the algorithm has already collected enough useful decisions in prior windows to offset the uninformative windows with low event counts.

\section{Discussion and Conclusions}\label{sec:discussion}

In this paper, we have introduced quantum analog asynchronous event-based graph neural network (QA-AEGNN), a novel framework that bridges three previously disjoint research areas: quantum computing with neutral-atom arrays, event-based neuromorphic vision, and graph neural networks. Our approach leverages the native Hamiltonian dynamics of Rydberg quantum processors to implement message-passing graph neural network computations directly in analog quantum hardware.

The key insight underlying QA-AEGNN is that the natural structure of neutral-atom quantum processors—with programmable atomic positions and distance-dependent interactions—provides a native substrate for encoding spatio-temporal event graphs. By mapping event coordinates to atomic positions such that geometric proximity reflects spatio-temporal adjacency, and by programming the Rydberg Hamiltonian to implement aggregation and combination operations through many-body interactions, we achieve a quantum implementation of event-driven GNNs \cite{schaefer2022aegnn} that exploits the massive parallelism and continuous dynamics of quantum systems.

Our experimental results on graph classification tasks and on neuromorphic data demonstrate that QA-AEGNN can outperform classical AEGNN. Consistently with prior art on quantum GNNs \cite{henry2021quantum}, the proposed quantum approach appears to better mix information across nodes of a graph, overcoming the limitations of classical message passing. These results suggest that quantum analog implementations may offer advantages beyond mere computational speedup, potentially extracting more informative features from graph-structured data through quantum interference and entanglement effects.

Several directions for future work emerge from this initial investigation:

\begin{itemize}
\item \textbf{Scaling to larger systems}: Our current experiments focus on relatively small graphs (up to 100 events). Investigating the scaling behavior of QA-AEGNN to larger event streams and higher-resolution event cameras will be essential for practical applications. Recent demonstrations of neutral-atom quantum processors with hundreds of qubits \cite{ebadi2022quantum} suggest that substantial scaling is achievable.

\item \textbf{Hybrid quantum-classical architectures}: Exploring more sophisticated combinations of quantum and classical processing, such as using quantum processors for early feature extraction followed by classical refinement, may offer practical advantages in near-term systems.

\item \textbf{Alternative quantum platforms}: While we have focused on neutral-atom implementations, other quantum computing platforms, such as superconducting qubits, trapped ions, and photonic systems, may offer complementary advantages for event-based graph neural networks. Comparative studies across platforms would be valuable.

\item \textbf{Theoretical analysis}: Developing rigorous theoretical frameworks for understanding the computational complexity and potential quantum advantages of QA-AEGNN, including formal characterizations of the feature spaces accessible to quantum versus classical implementations, remains an important open question.
\end{itemize}

\bibliographystyle{IEEEtran}
\bibliography{biblio}

\end{document}